\documentclass[
superscriptaddress,
 amsmath,amssymb,
 aps,
 prc,
 twocolumn,
floatfix
longbibliography
]{revtex4-2}

\usepackage[colorlinks,linkcolor=blue,anchorcolor=blue,citecolor=blue,urlcolor=blue]
{}
\usepackage{graphicx}
\usepackage{color}

\newcommand{\SU}[1]{\ensuremath{\mathrm{SU}( #1 )}}

\newcommand{\SpR}[1]{\ensuremath{\mathrm{Sp}( #1,\mathbb{R} )}}
\newcommand{\braket}[2]{\ensuremath{\langle #1 | #2 \rangle}}

\begin{document}

\date{\today}

\title{Comparing invariant-mass spectroscopy of $^8$B with \textit{ab initio} predictions}

\def\WUPHYS{Department of Physics, Washington University, St. Louis, Missouri 63130, USA.}
\def\FRIB{Facility for Rare Isotope Beams, Michigan State University, East Lansing, Michigan 48824, USA.}
\def\PAMSU{Department of Physics \& Astronomy, Michigan State University, East Lansing, Michigan 48824, USA.}

\def\Fudan{Key Laboratory of Nuclear Physics and Ion-beam Application (MOE), Institute of Modern Physics, Fudan University, Shanghai 200433, China.}
\def\Shanghai{Shanghai Research Center for Theoretical Nuclear Physics, NSFC and Fudan University, Shanghai 200438, China.}
\def\MSUPHYS{Department of Physics and Astronomy, Michigan State University, East Lansing, Michigan 48824, USA.}
\def\MSUCHEM{Department of Chemistry, Michigan State University, East Lansing, Michigan 48824, USA.}
\def\WUCHEM{Department of Chemistry, Washington University, St. Louis, Missouri 63130, USA.}
\def\ANL{Physics Division, Argonne National Laboratory, Argonne, IL 60439, USA.}
\def\WesternM{Department of Physics, Western Michigan University, Kalamazoo, Michigan 49008, USA.}
\def\Stores{Department of Physics, University of Connecticut, Storrs, Connecticut 06269, USA.}
\def\Lanzhou{Institute of Modern Physics, Chinese Academy of Sciences, Lanzhou 730000, China.}
\def\LSU{Department of Physics and Astronomy, Louisiana State University, Baton Rouge, Louisiana 70803, USA}

\author{R.J. Charity}
\affiliation{\WUCHEM}
\author{G.H. Sargsyan}
\affiliation{\FRIB}
\author{K.D. Launey}
\affiliation{\LSU}
\author {T.B. Webb}
\affiliation{\WUPHYS}
\author {K.W. Brown}
\affiliation{\FRIB}
\affiliation{\MSUCHEM}
\author {L.G. Sobotka}
\affiliation{\WUCHEM}
\affiliation{\WUPHYS}

\begin{abstract}
Levels in $^8$B have been investigated experimentally using the invariant-mass technique and compared with \textit{ab initio} calculations. Datasets obtained using $E/A$=69-MeV $^9$C and $^{13}$O beams on a Be target have been further analyzed to extend the level scheme of $^8$B for $E^*\lesssim$10~MeV. New levels were observed in the 2$p$+$^6$Li, $p$+$^3$He+$\alpha$, and the $p$+$^7$Be+$\gamma$ exit channels.
Momentum correlations between the decay fragments were also investigated in order to deduce the decay pathways and whether the decays are prompt or sequential. This nucleus and its mirror were also investigated in the \textit{ab initio} symmetry-adapted  no-core shell model. Correspondence between the newly observed and predicted levels were made based on the level energy and the decay modes. For positive-parity levels with $J\leq$3, all predicted levels can be connected to an experimental counterpart (as least tentatively) for $E^*\leq$8.4 MeV.

\end{abstract}

\maketitle

\section{INTRODUCTION}

Light nuclei like $^8$B and its mirror $^8$Li are ideal testing grounds for \textit{ab initio} nuclear
approaches as the calculations are computationally tractable. In these approaches, the nuclear force is modeled in the framework of the chiral effective-field theory (EFT) (e.g., \cite{Bedaque:2002,Epelbaum:2002,Etem:2003,Epelbaum:2006,Ekstrom:2013,Epelbaum:2014sza,RevModPhys.92.025004}). In particular, chiral potentials start from the degrees of freedom relevant to low-energy nuclear structure, that is, nucleons and pions, and account for the symmetry and symmetry-breaking patterns of the underlying theory of quantum chromodynamics.
Understanding low-lying states in these mirror nuclei is crucial for studies of isospin breaking effects, refining nuclear structure models, and probing chiral inter-nucleon interactions.

In this work, we expand the available level scheme of $^8$B using the invariant-mass technique applied to proton knockout reactions with a $^9$C beam. Some of the resonances observed in these knockout reactions, have also been verified using projectile fragmentation reactions with a $^{13}$O beam. 
In addition, we confront calculations in the \textit{ab initio} symmetry-adapted no-core shell model (SA-NCSM) \cite{Launey:2020, DytrychLDRWRBB20} using the NNLO$_{\rm opt}$ chiral potential \cite{Ekstrom:2013} with  previously known and newly observed states in $^8$B. The proton separation energy of $^8$B is very small (136~keV) and all excited states are resonances, i.e. particle unbound. This study then serves as a test of SA-NCSM calculations in the continuum, which are also benchmarked against energies obtained in the Green's function Monte Carlo (GFMC) approach~\cite{Wiringa:1995,Pieper:2008}.

In particular, the SA-NCSM, based on the concept of the no-core shell model (NCSM) \cite{NavratilVB00,BarrettNV13}, uses SU(3) symmetry to construct a complete basis, which reduces computational demands while retaining  nuclear structure information. An important consequence is that the SA-NCSM can reach ultra-large model spaces and hence, can accommodate large deformation and can couple to continuum degrees of freedom.
Furthermore, the SU(3) symmetry-adapted basis of SA-NCSM allows us to study the intrinsic deformation of the low-lying states in $^8$B, thereby informing about shape evolution and coexistence in $^8$B and $A=8$ isotopes. 

Comparisons of experimental and theoretical level properties allows us to assign spins for the new levels and account for all the predicted $^8$B levels up to 8.4~MeV. One of the new levels observed  is found to undergo a prompt 2$p$ decay. The nature of this 2$p$ emitter is the focus of a companion paper \cite{CharityShortArxiv}, where a comparison to the other known $A$=8 2$p$ emitters is made.

\section{DATASETS}
In this work we make use of three datasets all collected using the Si and CsI(Tl) telescopes from the HiRA apparatus \cite{Wallace:2007} and with secondary beams provided  by 
the Coupled Cyclotron Facility at the National Superconducting Cyclotron Laboratory at Michigan State University.  All three experiments utilized a 1-mm-thick Be target and 14 HiRA Si-CsI(Tl) telescopes were arranged in the same configuration 80 cm downstream from the target covering polar angles between 2$^\circ$ to 12$^\circ$.  In two of these datasets, $^8$B resonances are created from single-proton knockout reactions with an $E/A\approx$69-MeV $^9$C secondary beam.  The first of these, labeled  the $^9$C(1st) dataset, was optimized for the detection of $^8$C decay products  (protons and $\alpha$ particles) and signals from heavier elements saturated in the Si electronics \cite{Charity:2010,Charity:2011}. Thus this dataset is used only for the $p$+$^3$He+$\alpha$  exit channel.  The $^9$C beam for this dataset had an intensity of 1.6$\times$10$^5$ pps and a purity of $\approx$65\%.   In the second $^9$C dataset [$^9$C(2nd)] \cite{Brown:2014}, the electronics were upgraded to remove this limitation and the 2$p$+$^6$Li and $p$+$^7$Be exit channels were also examined. The $^9$C beam for this dataset was similar to the first set with an intensity of 1.2$\times$10$^5$ pps and purity of $\approx$52\%. However in this experiment, two of the 54 CsI(Tl) detectors did not operate. Unfortunately this included one of the innermost CsI(Tl) detectors, which reduced the efficiency and statistics for the $p$+$^3$He+$\alpha$ exit channel. Thus in the present work we show data from the $^9$C(1st) dataset for this exit channel and from the $^9$C(2nd) dataset for the other two channels. However, the results for the $p$+$^{3}$He+$\alpha$ exit channel are consistent in the two datasets.

Experimental parallel momentum distributions of $^8$B resonances from the $^9$C(2nd) dataset have been compared with theoretical models  of direct proton knockout in Refs.~\cite{Bonaccorso:2018,Charity:2020} and the higher-momenta regions are consistent with this process. At lower momenta there is need for additional yield from presumably more complex multi-step processes. As one expects the ground state of $^9$C ($J^\pi =3/2^-$) to be largely $p$-shell, direct knockout should involve the removal of either a $p_{1/2}$ or $p_{3/2}$ proton producing states of spin 0$^+$, 1$^+$, 2$^+$, or 3$^+$ in $^8$B.  However, given the neighboring $^{9,10,11}$N isotopes have $s$-wave ground states \cite{ENSDF,Hooker:2017,Charity:2023b}, it is possible there maybe  some level of the proton intruder $(1s_{1/2})^2$ configuration in the ground state of $^9$C as well.  Knockout of an $s_{1/2}$ proton would then produce 1$^-$ or 2$^-$ states.

In order to confirm some of the new resonances observed in the knockout data, we have also examined a dataset obtained with an $E/A$=69.5-MeV beam of $^{13}$O of intensity of 3.7$\times$10$^5$ pps and purity 80\% \cite{Webb:2019,Webb:2020,Charity:2021}. Here $^8$B resonances are created by projectile fragmentation with a net loss of five  nucleons from the projectile.   Such a reaction may not be as restrictive as to the $J^\pi$ of the resonances created, but in any case this dataset did not yield evidence for any  further $^8$B resonances. However, it did confirm the existence of some of the new resonances found in the one-proton knockout datasets.  For the $^9$C(2nd) dataset, the experiment was performed with the CAESAR CsI(Na) $\gamma$-ray detectors \cite{Weissharr:2010} surrounding the target which provided information on coincident $\gamma$ rays.

\section{RESULTS}
\subsection{2$p$+$^{6}$Li channel}

For the 2$p$+$^6$Li exit channel, one must consider two possibilities for the $^6$Li fragment. This fragment could be created in its 1$^+$ ground state or in its 0$^+$ second excited state (isobaric analog state) which decays by emitting a 3.562-MeV $\gamma$ ray.  All other $^6$Li excited states decay predominantly by particle emission. The excitation energy calculated from the invariant mass assuming the $^6$Li fragment was in its ground state, i.e. no coincident $\gamma$ ray, is designated $E^*_{n\gamma}$. If the $^6$Li fragment was formed in its 0$^+$ excited state, then the real excitation energy will be larger by the $\gamma$-ray energy (3.562~MeV).  
\begin{figure}[!htb]
\includegraphics[width=0.9\linewidth]{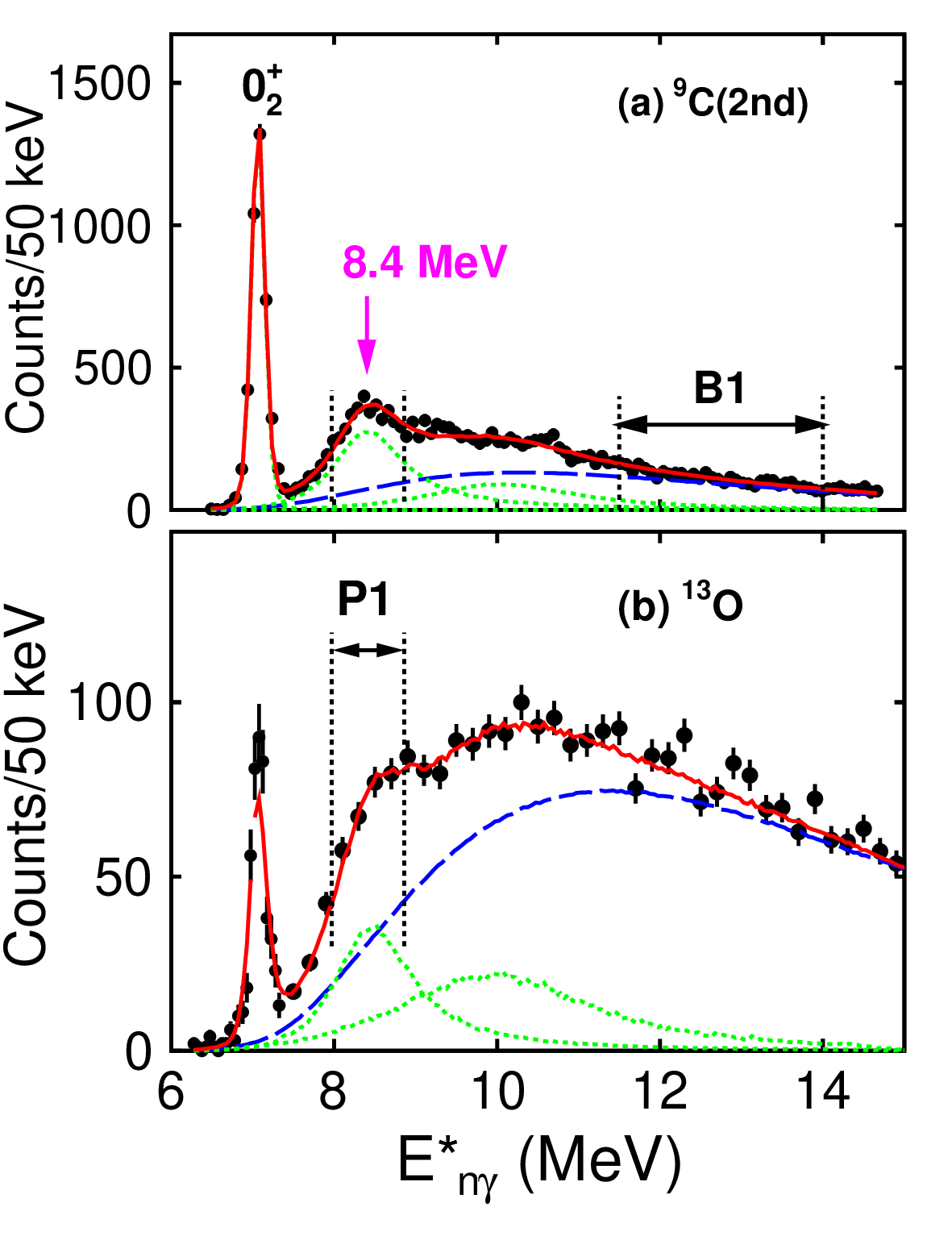}
\caption{Excitation energy in $^8$B obtained from the invariant-mass of detected 2$p$+$^6$Li events assuming the $^6$Li fragment was in its ground state. Data in (a) and (b) were obtained with the $^9$C(2nd) and $^{13}$O datasets, respectively. The red curves show fits to these data with  contributions from resonance peaks given by the dotted green curves and the background shown as the dashed blue curves. The gate $P1$ about the 8.4-MeV state is used for the correlations shown Fig.~\ref{fig:Corr_2pLi6compare}.  
} 
\label{fig:Inv_2pLi6}
\end{figure}

The distributions of $E^*_{n\gamma}$ obtained from  the $^{9}$C(2nd) and $^{13}$O datasets are displayed in Fig.~\ref{fig:Inv_2pLi6}.   The narrow peak observed in both datasets at $E^*_{n\gamma}\approx$ 7.1~MeV is associated with the 0$^+_2$ isobaric analog state in  $^8$B that undergoes prompt 2$p$ emission to the 0$^+$ isobaric analog state in $^6$Li. As there is an accompanying $\gamma$ ray \cite{Brown:2014}, its real excitation energy is 10.619 MeV \cite{ENSDF}. The 2$p$ decay of this state from the  $^9$C(2nd) dataset has been discussed in Ref.~\cite{Brown:2014}.  In Fig.~\ref{fig:Inv_2pLi6}(a)  there is a peak at $E^*_{n\gamma}$=8.4~MeV and excess yield is also confirmed in the results from the $^{13}$O dataset in 
Fig.~\ref{fig:Inv_2pLi6}(a) where a shoulder feature is present.

\begin{figure}[!htb]
\includegraphics[width=1.\linewidth]{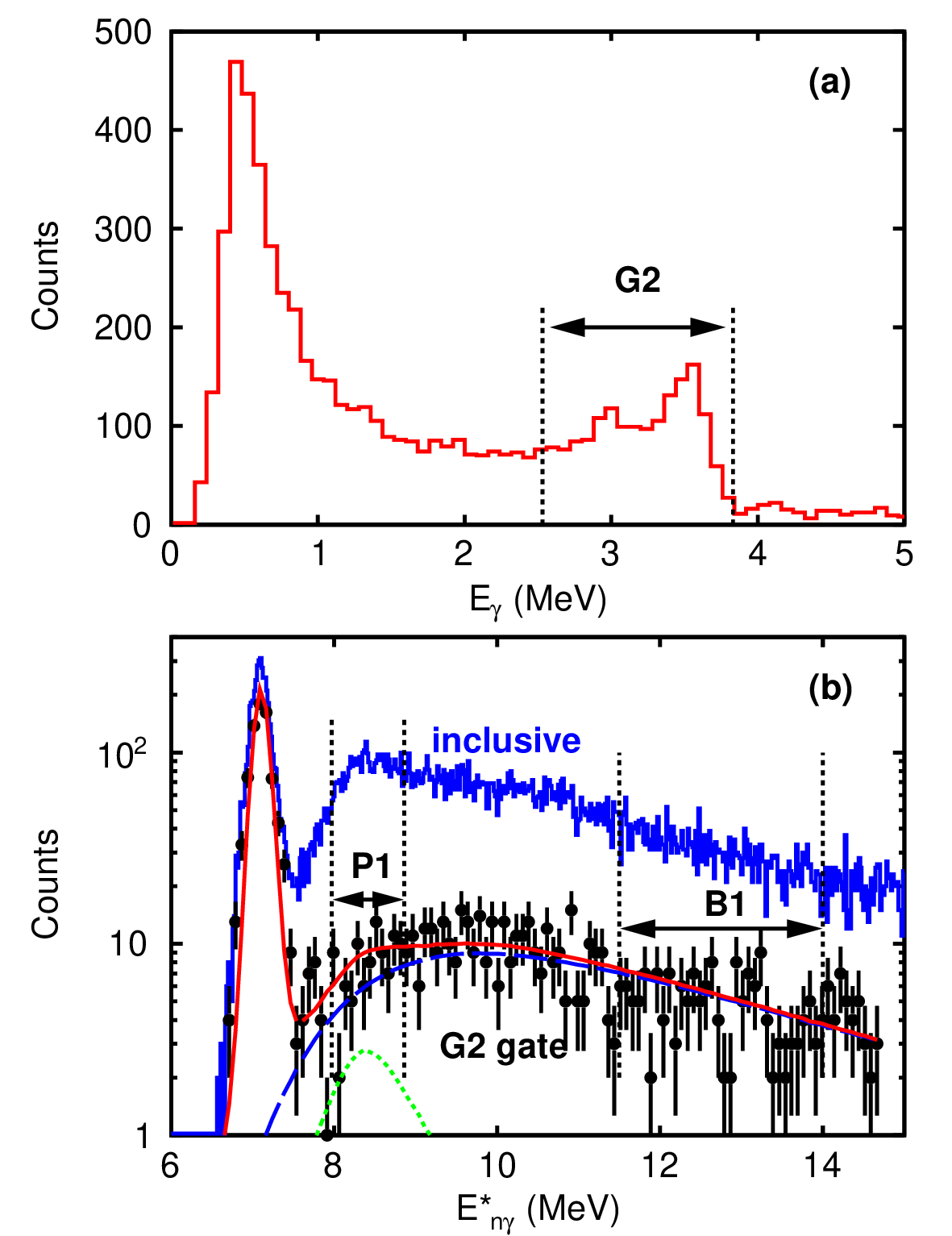}
\caption{(a) Energy spectrum of $\gamma$-rays detected in coincidence with the 2$p$+$^6$Li channel in the $^9$C(2nd) dataset.  The gate $G2$ around the 3.55-MeV full-absorption peak and its first escape peak is used to enhance the yield associated with the 0$^+$ state relative to the 1$^+$ ground state of  $^6$Li. (b)
$E^*_{n\gamma}$ distribution obtained from the invariant mass of 2$p$+$^6$Li events.
The blue histogram is for all events while the black data points have a $\gamma$ ray in the $G2$ gate. A fit to the latter with the same contributions as in Fig.~\ref{fig:Inv_2pLi6} is shown by the curves.
}
 \label{fig:gamma}
\end{figure}

\begin{figure}[!htb]
\includegraphics[width=1.\linewidth]{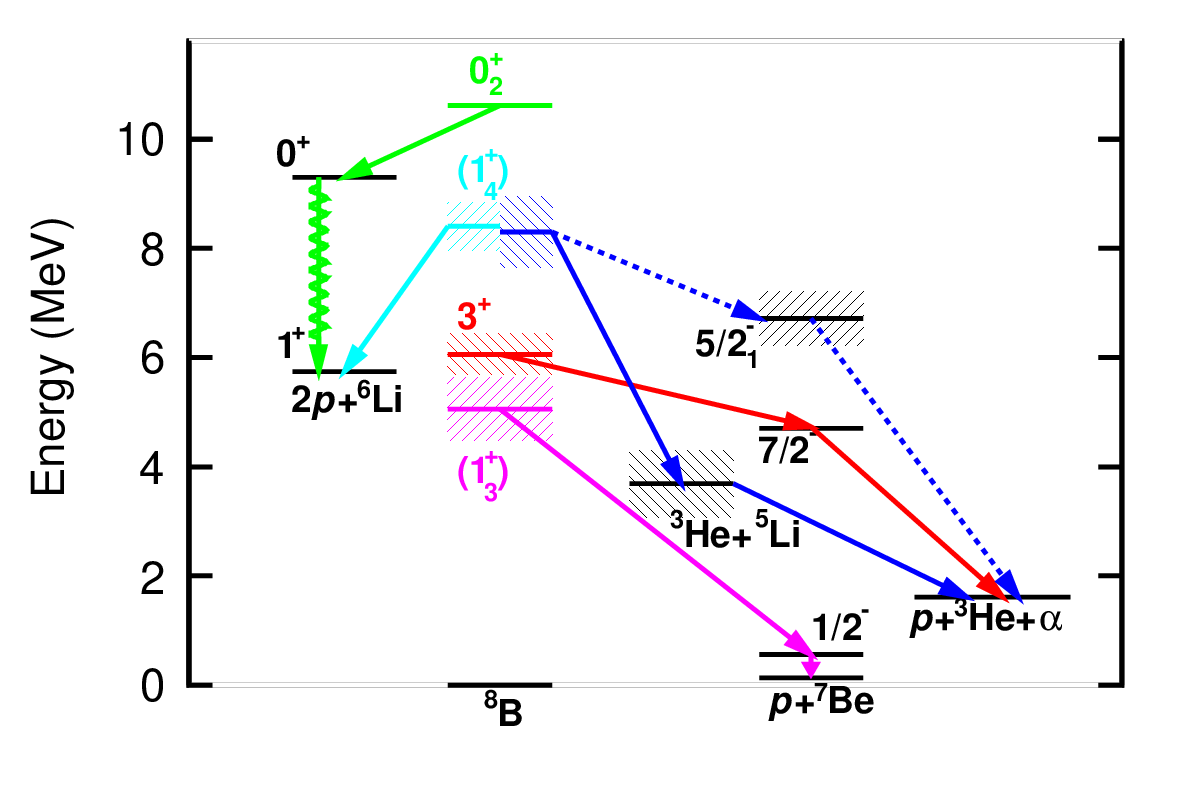}
\caption{Level diagram of the states observed in the  2$p$+$^{6}$Li, $p$+$^{3}$He+$\alpha$, 
and $p$+$^7$Be+$\gamma$ channels where the indicated sequential decay pathways have been observed.  The 0$^+_2$ state in $^8$B decays by 2$p$ emission to the excited 0$^+$ state in $^6$Li which subsequently decays to the ground state via $\gamma$-ray emission. This 0$^+_2$ state corresponds to the narrow peaks in Figs.~\ref{fig:Inv_2pLi6}(a) and \ref{fig:Inv_2pLi6}(b).}
 
\label{fig:levelpaHe3}
\end{figure}

For the former $^9$C(2nd) dataset, the CAESAR $\gamma$-ray array \cite{Weissharr:2010} was deployed along with HiRA allowing one to determine whether these 2$p$+$^6$Li invariant peaks are associated with the 1$^+$ ground or 0$^+$ isobaric analog state in $^6$Li. The full-absorption and first-escape peak associated with this $\gamma$ ray are clearly seen in the coincidence $\gamma$-ray-energy distribution shown in Fig.~\ref{fig:gamma}(a). Using the gate $G2$ around these peaks, the coincidence $E^*_{n\gamma}$ spectrum is shown as the data points in Fig.~\ref{fig:gamma}(b) which can be compared with the ungated (blue histogram) version. The 7.1-MeV peak is present in the gated spectrum and thus is coincident with the 3.563-MeV $\gamma$ ray. Thus its true excitation energy is 3.563-MeV higher. The 8.4-MeV peak can be identified in inclusive spectrum, but is absent from the gated spectrum indicating that this peak is associated with decay to the ground state of $^6$Li. The two decay paths are illustrated in the level diagram of Fig.~\ref{fig:levelpaHe3}.

The red curves in Figs.~\ref{fig:Inv_2pLi6}(a) and \ref{fig:Inv_2pLi6}(b) are fits to the total spectra with the contributions from the individual peaks (dotted green curves) and the background (dashed blue curves) displayed.  Apart from the two peaks already mentioned, a third broad peak centered at $\approx$10~MeV is required to fit the  spectra. This third peak is most prominent in the $^{13}$O dataset, but its presence greatly improves the fits in both datasets.

In these fits, the shape of the background is determined using weighted event mixing (Appendix~\ref{sec:mix2pLi6}), but its magnitude is a fit parameter. The centroid and width for the narrow 0$^+_2$ state were fixed to be consistent with the ENSDF value \cite{ENSDF} and a Breit-Wigner intrinsic line shape is assumed. For the other two wider peaks, which may have asymmetric line shapes, we have used profiles consistent with diproton emission in the $R$-matrix approximation \cite{Barker:2001}, but for simplicity the two protons are assumed to have zero relative energy.  The effects of the detector efficiency and resolution in this and other fits were included using Monte Carlo simulations as in Ref.~\cite{Charity:2019}.  The fitted parameters for these peaks and their uncertainties are listed in Table~\ref{tbl:2pLi6}.

\begin{table}[ht]
\caption{\label{tbl:2pLi6} Level parameters obtained from fitting the $E^*_{n\gamma}$ distributions in Fig.~\ref{fig:Inv_2pLi6}. The first (lowest-energy) peak was fit with its known $E^*_{n\gamma}$ value \cite{ENSDF}. The second and third peak refer to the next two peaks ordered by increasing $E^*_{n\gamma}$. 
}
\begin{ruledtabular}
\begin{tabular}{ccccc}

 dataset&  \multicolumn{2}{c}{second peak} &   \multicolumn{2}{c}{third peak}   \\
&  $E^*$ &  $\Gamma$ & $E^*$ & $\Gamma$ \\
 &             (MeV)   &    (MeV) & (MeV) & (MeV) \\
\colrule
$^9$C & 8.38(6)  & 0.94(10) & 10.02(36) &  1.87(64) \\
$^{13}$O    & 8.41(5)   & 0.82(10) & 10.16(37) &  3.70(95)  

\end{tabular}
\end{ruledtabular}
\end{table}

\subsection{ $p$+$^{3}$He+$\alpha$ channel}
Evidence for a possible second decay branch of the 8.4-MeV state is obtained in the $p$+$^3$He+$\alpha$ exit channel from both $^9$C datasets.  There are a number of different sequential pathways involving $^7$Be and $^5$Li intermediate states that could give rise to this channel. The level diagram in Fig.~\ref{fig:levelpaHe3}
summarizes the decay pathways with the identified intermediate states.

\begin{figure}[!htb]
\includegraphics[width=1.\linewidth]{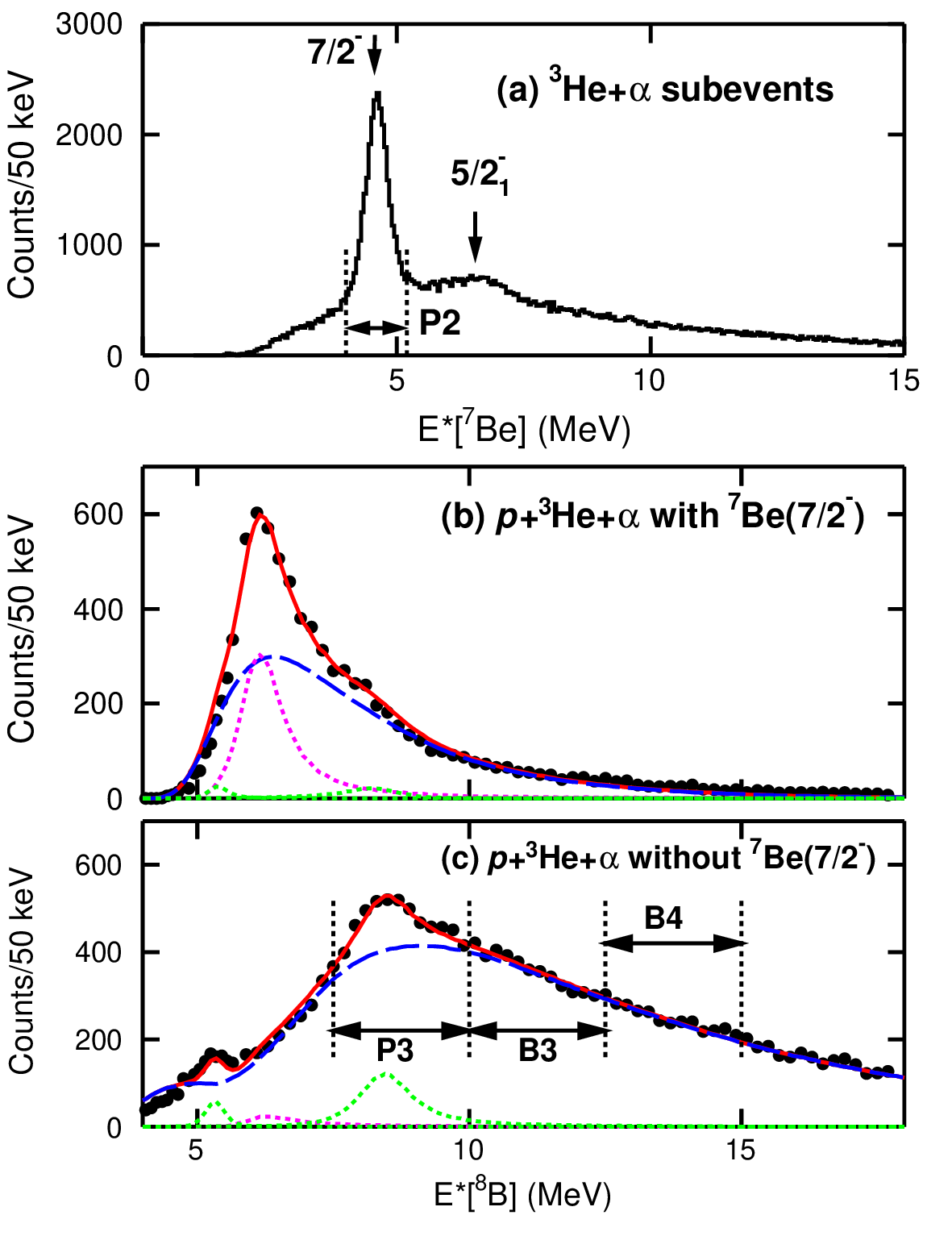}
\caption{Spectra obtained from $p$+$^{3}$He+$\alpha$ events produced with the  $^9$C(2nd) dataset. (a)  The $^7$Be excitation-energy distribution obtained from the $^3$He+$\alpha$ subevents. The observed peaks are labeled by their known $J^\pi$.  The gate $P2$ is used to select or veto the 7/2$^-$ intermediate state in $^7$Be. The $^8$B excitation energy spectrum from the $p$+$^3$He+$\alpha$ events obtained in coincidence with this gate is shown in (b), while (c) is obtained by vetoing with this gate. The solid red curves are fits to the data with  peak contributions shown by the dotted green curves and backgrounds indicated by the dashed blue curves.
The gate $P3$ around the 8.2-MeV peak is used in the constructions of the correlations shown in Fig.~\ref{fig:corr_pahe3}.
}
\label{fig:Inv_pahe3}
\end{figure}

The distribution of $^7$Be excitation energy obtained from the $^3$He+$\alpha$ subevents in the $^9$C(1st) dataset is plotted in  Fig.~\ref{fig:Inv_pahe3}(a). The 7/2$^-$ resonance and a smaller contribution from the 5/2$^-_1$ state are evident in this distribution. 
Given the strong yield of the $^7$Be(7/2$^-$) resonance, we have subdivided the events into those inside and outside of the gate $P2$ which selects this state as an intermediate in the decay. The deduced $^8$B excitation-energy spectra for these two groups are displayed in Figs.~\ref{fig:Inv_pahe3}(b) and \ref{fig:Inv_pahe3}(c), respectively.   The spectrum associated with the 7/2$^-$ intermediate state, Fig.~\ref{fig:Inv_pahe3}(b), is dominated by a peak at $\approx$6.1 MeV which was first described in Ref.~\cite{Charity:2011} where the correlations measured between the two sequential decay axes were found consistent with either  $J^\pi$=3$^\pm$ or 4$^{\pm}$. As 3$^-$ and 4$^\pm$ states cannot be populated by knocking out either a $p$-wave or a $s$-wave proton from $^9$C, this state must have  $J^\pi$=3$^+$.  

The spectrum generated by vetoing the 7/2$^-$ intermediate state, Fig.~\ref{fig:Inv_pahe3}(c), shows the presence of a state at $\approx$8.2~MeV and, with much lower yield, another state at $\approx$5.4-MeV peak. The red curves in Figs.~\ref{fig:Inv_pahe3}(b) and \ref{fig:Inv_pahe3}(c) are joint fits to these experimental spectra where the shape of the background contribution is discussed in Appendix~\ref{sec:mixpHe3a}. The intrinsic line shapes are from $R$-matrix theory for sequential decay of isolated resonances \cite{Kalanee:2013}. The 6.1-MeV state, see Fig.~\ref{fig:Inv_pahe3}(b), is taken to decay by $p$-wave proton emission to the $^{7}$Be(7/2$^-$) resonance whose intrinsic line shape is taken from Ref.~\cite{Charity:2025}.  The other states were treated as $^{3}$He decays through a $^5$Li(g.s.) intermediate (see later). The intrinsic line shape of $^5$Li(g.s.) is taken from the $R$-matrix parameters of \cite{Woods:1988}.

The peak parameters from the fit are listed in Table~\ref{tbl:paHe3} as well as results from fitting the other two datasets. The $\approx$5.4 and $\approx$8.2-MeV states sit on large backgrounds and the fitted widths are very sensitive to the shape of the background distribution assumed in the fit (discussed in Appendix~\ref{sec:mixpHe3a}) and hence have large uncertainties. In addition to this contribution, the quoted uncertainties also include contributions from the uncertainties in the energy calibrations and the statistical error in the fitting process. There is no evidence for the $\approx$5.4 and $\approx$8.2-MeV peaks in the $^{13}$O dataset. 

It is possible the 8.2~MeV state in the $p$+$^3$He+$\alpha$ channel and the 8.4-MeV peak in the 2$p$+$^6$Li channel are branches on the same state because their centroids and widths are quite similar (see Tables \ref{tbl:2pLi6} and \ref{tbl:paHe3}).  The lack of an observation of the $\approx$8.2-MeV peak in the $p$+$^3$He+$\alpha$ channel from the $^{13}$O dataset is not significant because this resonance would be difficult to discern above a significantly increased background. Relative to the expected yield for this resonance, the background in the $p$+$^3$He+$\alpha$ channel has increased by a factor of 50 in the $^{13}$O dataset compared with the $^9$C datasets.

\begin{figure}[!htb]
\includegraphics[width=1.\linewidth]{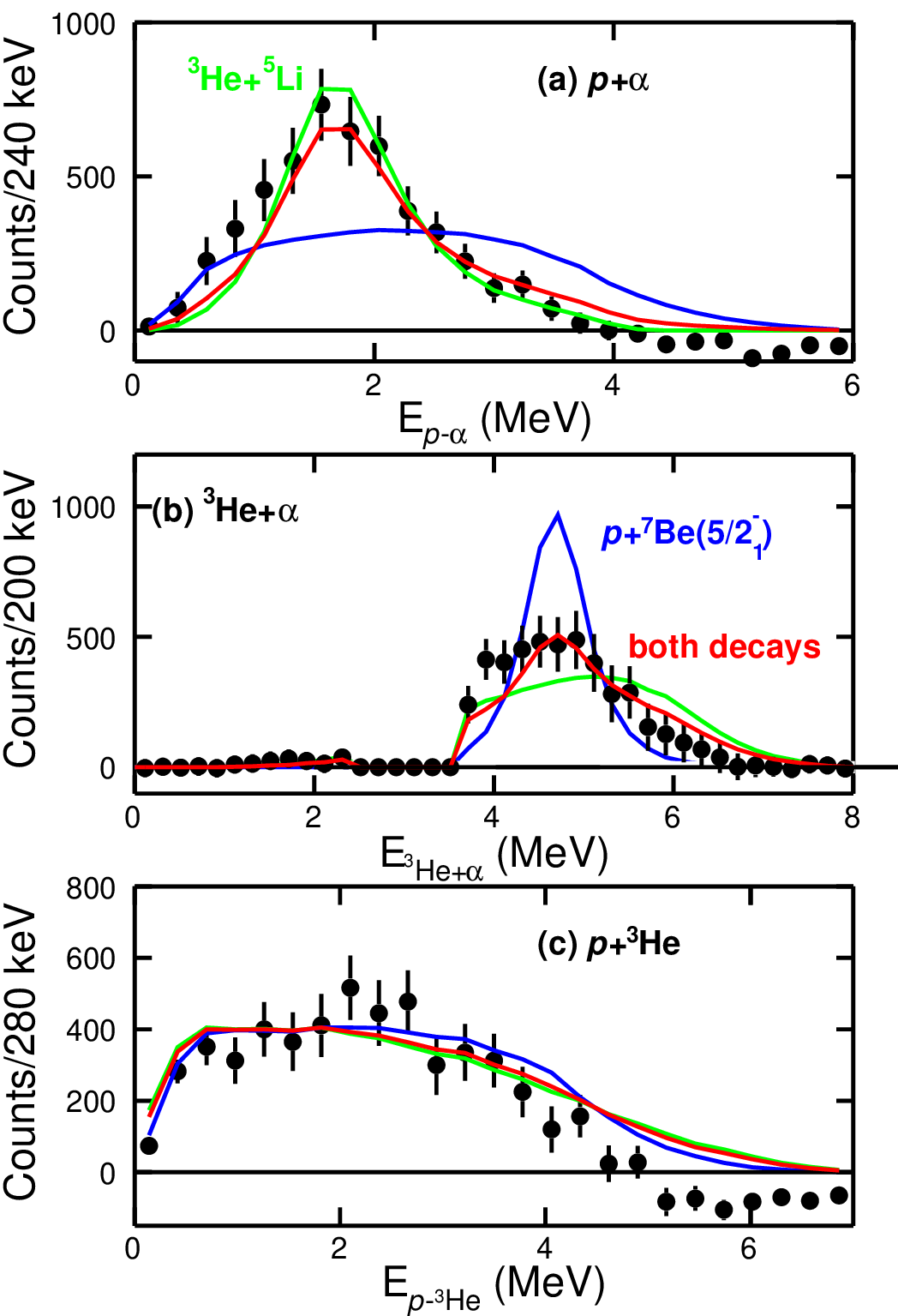}
\caption{Data points correspond to the background-subtracted correlations in the  decay of the 8.2-MeV state in $^8$B to the $p$+$^3$He+$\alpha$ channel obtained with the $^{9}$C(1st) dataset. The projections of the correlations on the (a) $p$-$\alpha$, (b) $^3$He-$\alpha$, and (c) $p$-$^3$He relative-energy axes are shown. The solid green and blue curves shown predictions of sequential decay simulations through the $^5$Li(g.s.)  and $^7$Be(5/2$^{-}_{1}$) intermediate states, respectively. The solid red curve is the best fit allowing contributions from both sequential decay scenarios.
}
 \label{fig:corr_pahe3}
\end{figure}

To infer the  nature of the decay mechanism populating the $p$+$^3$He+$\alpha$ channel, we have examined the correlations between the momentum of the three decay products for the gate $P3$ around the 8.2-MeV peak in Fig.~\ref{fig:Inv_pahe3}. In the fits, the peak yield in this gate represent 20(4)\% of the total and thus great care must be made in subtracting the background correlations. The correlations in the background events and their dependence on $E^*$ have been examined extensively, and discussed in Appendix~\ref{sec:mixpHe3a}, for excitation energies above this gate which is almost entirely background in our fits. Correlations for the background events in the $P3$ gate are then estimated by extrapolation down in energy. The background-subtracted correlations, expressed as the spectra of the relative energies between each fragment type, are plotted in Fig.~\ref{fig:corr_pahe3}.  Note that the $P3$ gate is applied to the $P2$-vetoed spectrum, i.e. the events in Fig.~\ref{fig:Inv_pahe3}(c), and thus the region of the $^{3}$He-$\alpha$ relative-energy spectrum around the $^{7}$Be(7/2$^-$) peak ($P2$ gate) is missing. It is clear that our extrapolation of the background correlations is not perfect because the background-subtracted spectra in Figs.~\ref{fig:corr_pahe3}(a) and \ref{fig:corr_pahe3}(c) go slightly negative at high energies. Nevertheless, gross features of the decay mechanism can be inferred.

The distribution of the $p$-$\alpha$ relative energy in Fig.~\ref{fig:corr_pahe3}(a) is dominated by a wide peak at $\approx$1.7~MeV consistent with the presence of the broad $^5$Li ground state. The green curves in all three panels of Fig.~\ref{fig:corr_pahe3} are from a simulation of sequential decay via the $^8$B$\rightarrow^{3}$He+$^5$Li(g.s.)$\rightarrow p$+$^{3}$He+$\alpha$ pathway.   These curves reproduce the experimental spectra reasonably well, except in the nonphysical regions where the experimental values are negative. An alternative sequential decay scenario could be $^8$B$\rightarrow p$+$^7$Be(5/2$^-_1$)$\rightarrow p$+$^3$He+$\alpha$, a scenario indicated by the blue curves in Fig.~\ref{fig:corr_pahe3}. In this second scenario, the $^7$Be(5/2$^-_1$) intermediate state would show up as a peak in the $^{3}$He+$\alpha$ relative energy spectrum shown in Fig.~\ref{fig:corr_pahe3}(b)  at $\approx$ 4.7~MeV. While the experimental data do have a broad peak close to  $E_{^{3}\rm{He}+\alpha}\approx$4.7~MeV, it is  smaller and wider than the simulated distribution (blue curve). This simulation makes use of new more-precise measurements of the 
centroid and width of the 5/2$^-_1$ state \cite{Charity:2025} where the centroid is $\approx$300~keV lower than the ENSDF value and the width is roughly half of the ENSDF width.

A better fit (red curves) to the correlations can be obtained by allowing contributions from both  scenarios.  A joint fit was made to the $E_{p+\alpha}$ and $E_{^{3}\rm{He}+\alpha}$ data, the E$_{p+^{3}\rm{He}}$ data shown in Fig.~\ref{fig:corr_pahe3}(c) does not discriminate between these two branches.  The fitted branching ratios are 73(7)\% and 27(7)\% for the $^5$Li and $^7$Be(5/2$^-_1$) intermediate states, respectively.  Due to the problems with the background subtraction, the quoted statistical uncertainties in these branching ratios may be too small. Nevertheless, the conclusion that the dominant decay mode of the 8.2~MeV state is through $^{3}$He+$^5$Li is robust.

\begin{table*}[ht]
\caption{\label{tbl:paHe3} Level parameters obtained from fitting the $p$+$^3$He+$\alpha$ excitation-energy distributions. Three peaks are listed in order of increasing energy.}
\begin{ruledtabular}
\begin{tabular}{ccccccc}

 dataset&  \multicolumn{2}{c}{1st peak} &   \multicolumn{2}{c}{2nd peak} & \multicolumn{2}{c}{3rd peak}   \\
&  $E^*$ &  $\Gamma$ & $E^*$ & $\Gamma$ & $E^*$ & $\Gamma$\\
 &             (MeV)   &    (MeV) & (MeV) & (MeV) &(MeV) &(MeV)\\
\colrule

$^9$C(1st) & 5.33(11)  & 0.35(61) & 6.104(24) &  0.90(25) & 8.26(26) & 1.11(55)\\
$^9$C(2nd) & 5.43(15)  & 0.42(44) & 6.016(92) & 0.75(20) & 8.20(26) & 1.47(42)\\

$^{13}$O    &            &  & 6.022(44) &  0.94(9) &              \\  
\end{tabular}
\end{ruledtabular}
\end{table*}

\subsection{$p$+$^7$Be channel}
\label{sec:pBe7}
A presentation of the $p$+$^7$Be decay channel from the $^9$C(2nd) dataset was previously made in Ref.~\cite{Brown:2017}. In this section we extend the analysis of these data and derive limits for the branching ratio of the 8.4-MeV state to this channel.  The 1/2$^-$ first excited state of $^7$Be ($E^*$=0.429~MeV) decays by $\gamma$-ray emission.  Thus a $^8$B parent state that decays to both the ground and this first excited state would contribute two peaks to the invariant-mass spectrum separated from each other by 0.429 MeV. 

\begin{figure}[!htb]
\includegraphics[width=1.\linewidth]{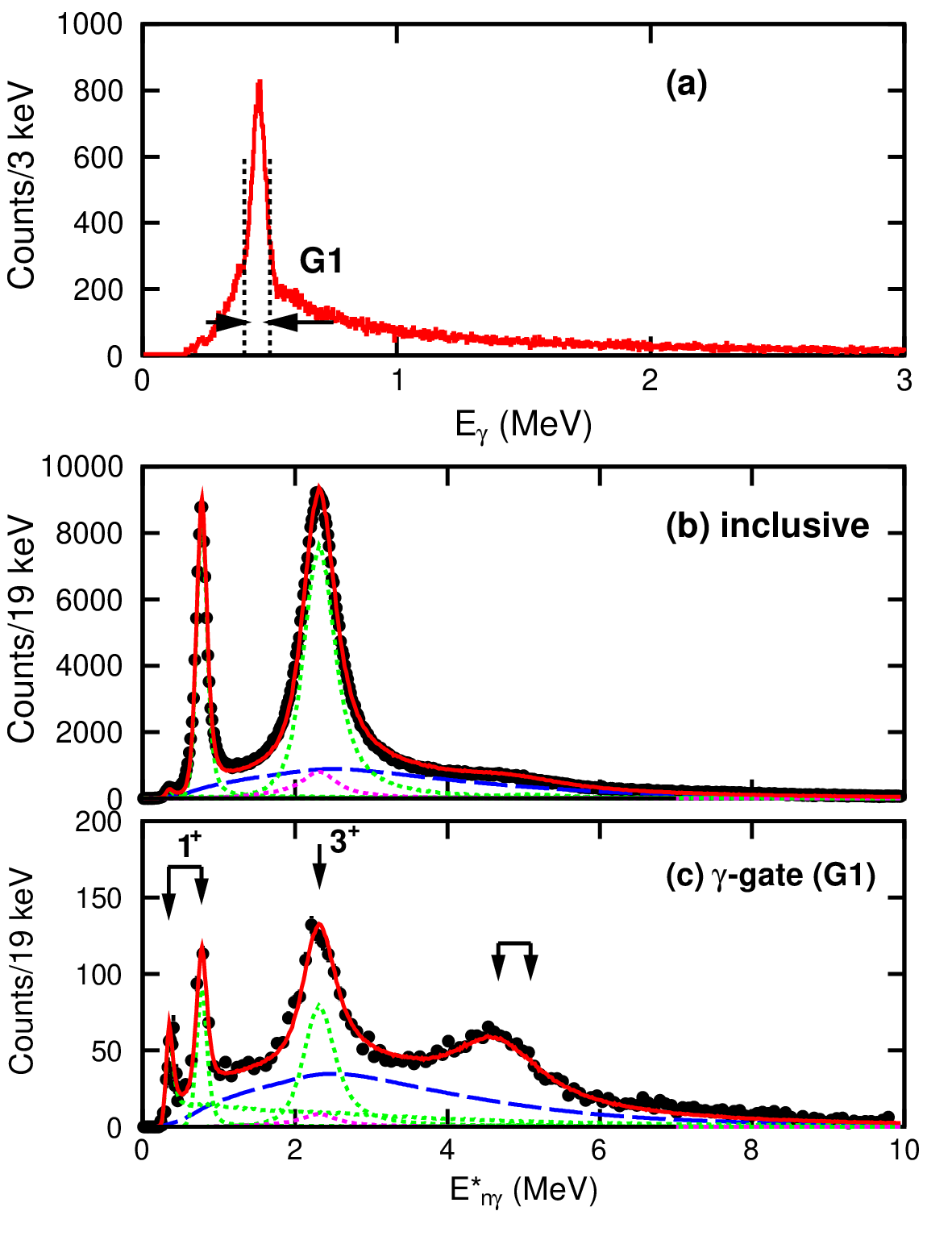}
\caption{(a) Energy spectrum of $\gamma$ rays detected in coincidence with the $p$+$^7$Be channel.  (b) Inclusive spectrum of $E^*_{n\gamma}$ obtained from the invariant mass of $p$+$^7$Be events. (c) Spectrum of the same quantity, but in coincidence with a $\gamma$ ray in the $G1$ gate. Red curves in (b) and (c) are from joint fits with peak (dotted green curves) and background (dashed blue curves) components indicated.  The levels discussed in the text that can decay to both the ground and first excited states of $^7$Be are marked by the paired arrows. The higher-energy (lower-energy) arrow corresponds to the $E^*_{n\gamma}$ value for proton decay to ground  (first excited) state of $^7$Be.
}
 \label{fig:Inv_pBe7}
\end{figure}

The $\gamma$-ray energy spectrum measured in coincidence with the $p$+$^7$Be channel, displayed in Fig.~\ref{fig:Inv_pBe7}(a), shows the presence of 0.429-MeV $\gamma$ rays from the decay of the 1/2$^-$ first excited state sitting on a significant background. A gate $G1$ around this peak is used to enhance proton decays to the 1/2$^-$ state as compared with the ground state. Inclusive and $G1$-gated $E^*_{n\gamma}$ spectra are shown in Fig.~\ref{fig:Inv_pBe7}(b) and \ref{fig:Inv_pBe7}(c), respectively.   The $G1$-gated spectrum shows the enhancement of two peaks, the narrow lowest-energy peak and the broad peak at $E^*_{n\gamma}\approx$4.7~MeV. The former is a small decay branch of the 1$^+$ first excited state of $^8$B \cite{Brown:2017} whose presence is barely visible in the inclusive spectrum. The main proton decay branch of this level is to the $^7$Be(g.s.) which corresponds to the peak just higher in energy seen in Fig.~\ref{fig:Inv_pBe7}(c). The 3$^+$ second excited state of $^8$B is evident by the peak at $E^*_{n\gamma}$=2.3~MeV. Assuming $p$-wave proton emission, this state should not decay to the $^7$Be(1/2$^-$) state and its greatly suppressed presence in the $G1$-gated spectrum is presumably due the background contribution in the $\gamma$-ray gate.
The broad state at $E^*_{n\gamma}\approx$4.7 MeV, which is only clearly seen in the $G1$-gated spectrum, corresponds to a previously unknown $^8$B state.

The red curves in Figs.~\ref{fig:Inv_pBe7}(b) and \ref{fig:Inv_pBe7}(c) are from joint fits to the inclusive and $G1$-gated spectra. The lineshapes of the 1$^+$ and newly observed state at $E^*_{n\gamma}\approx$4.7 MeV are each treated in the $R$-matrix approximation as isolated resonances with two channels. The magnitude, centroid, and partial width of each channel are varied.  These peaks sit upon background distributions (dashed blue curves) generated by the weighted event mixing procedure \cite{Charity:2023}. In addition there is a small fixed contribution (dotted magenta curves) from the 2$p$ decay of the 5.75-MeV level in $^9$C that sequentially decays through the $^8$B(3$^+$) intermediate state \cite{Brown:2017}. This component was determined using a fit of the 2$p$+$^7$Be data and simulations of the detector response to fix the contribution when only one of the two protons is detected.

Following the procedure used in previous joint fits of invariant-mass data \cite{Charity:2022}, we assume the background in the $G1$ gate gives rise to a scaled down copy of the inclusive peak yields plus, for the $p$+$^7$Be(1/2$^-$) branches, extra yield fixed by the inclusive yield times the $\gamma$-ray detection efficiency. 
The $p$+$^7$Be(1/2$^-$) branch of the new level constrains its centroid to  $E^*_{n\gamma}$=4.666(20)~MeV, corresponding to an excitation of 5.095(20)~MeV. Its fitted width is $\Gamma$=1.15(7) MeV. The branching ratio of this level to $^7$Be(g.s.) is only restricted to $\lesssim$50\% of the 1/2$^-$ branch.  
This state is at the same excitation as a 1$^-$ state reported in two studies, $E^*=5.0(4)$~MeV \cite{Yamaguchi:2009} and 5.1~MeV \cite{Mitchell:2010}, A state of this $J^\pi$ could be produced by the knockout of a $1s_{1/2}$ intruder proton in $^9$C. 
However, our fitted width is between, and inconsistent with, the reported values of 150(100)~keV and 4.6~MeV from \cite{Yamaguchi:2009} and \cite{Mitchell:2010}, respectively.  

No indication of peaks associated with 1$p$ decay branches of the 8.4-MeV $^8$B state can be found in these $E^*_{n\gamma}$ spectra. However, our sensitivity is poor. Such branches will have low detection efficiency as the $\approx$8-MeV protons will most often be emitted to laboratory angles outside the acceptance of the apparatus. Only situations where the proton is emitted in the  forward or backward directions can they be detected and such events have poor reconstructed invariant-mass resolution \cite{Charity:2019}. Nevertheless, we had added additional peaks in the fits to the $E^*_{n\gamma}$ distribution associated with 1$p$ decay to the ground and first excited states of $^7$Be. Correcting for the detection efficiencies of the different branches, the maximum possible yield from the two 1$p$ branches consistent with the experimental data is 2.2 times that found for the 2$p$ branch. In fitting the $G1$-gated spectrum alone, the limit for the $p$+$^7$Be(1/2$^-$) branch is 1.3 times the 2$p$ branch. Ignoring the possible contribution from a $p$+$^3$He+$\alpha$ branch, the 2$p$ branching ratio is a least 30\%. This reduces to at least 23\% if the $p$+$^3$He+$\alpha$ branch is included. Without ambiguity, the 2$p$ branch is a significant deexcitation pathway.

\section{Theory}

The \textit{ab initio} SA-NCSM is based on the no-core shell-model concept, but formulated with an \SU{3}-coupled or \SpR{3}-coupled symmetry-adapted basis (see, e.g., Refs. \cite{LauneyDD16,DytrychLDRWRBB20} and references therein). This enables physically relevant selections of the SA-NCSM model spaces, which in turn allows the nucleons to occupy much larger spaces. This is essential to accommodate large deformation, clustering and coupling to the continuum. In the SA-NCSM with continuum \cite{LauneyMD_ARNPS21}, the interior wave function is obtained from high-quality SA-NCSM calculations, whereas the asymptotics in the exterior can be straightforwardly recovered by using the exact Coulomb wavefunctions at large distances through, e.g., a microscopic R-matrix technique (currently implemented for partitionings of two clusters). In this way, the SA-NCSM framework can be utilized to provide structure observables, including $B(E2)$ transition strengths for Mg isotopes \cite{Ruotsalainen19,launeyPPNP2026} and beta decays \cite{Sargsyan_A8}, and in addition, reaction observables, such as partial widths and asymptotic normalization coefficients \cite{DreyfussLESBDD20,Sargsyan_A8,SargsyanLSMD2023}, response functions and sum rules \cite{BurrowsEWLMNP19, Baker:2024}, as well as alpha capture and knock-out reaction cross sections \cite{DreyfussLESBDD20,Sargsyan:2025}.

Similar to the conventional NCSM \cite{NavratilVB00,NavratilVB00b}, the SA-NCSM employs the single-particle basis of a harmonic oscillator (HO) potential, where major shells are separated by the parameter $\hbar \Omega$. The many-body model space is capped at an $N_{\rm max}$ cutoff, which defines the maximum total number of oscillator quanta above the valence-shell configuration. As $N_{\rm max}$ approaches infinity, the results converge toward the exact values and become independent of the HO parameter $\hbar \Omega$. The SA-NCSM utilizes a non-relativistic nuclear Hamiltonian, which includes translationally invariant inter-nucleon interaction alongside the Coulomb interaction between protons. 
SA-NCSM computes eigenvalues and eigenvectors of the nuclear Hamiltonian, which are subsequently used to evaluate nuclear observables. Since calculations are performed in laboratory coordinates, spurious center-of-mass excitations are removed from the low-lying spectrum via a Lawson term \cite{Lawson74,DytrychMLDVCLCS11}.  Within a given complete $N_{\rm max}$ model space, the SA-NCSM results are equivalent to those obtained with the NCSM for the same interaction. 

The many-nucleon basis states of the SA-NCSM provide important information, namely, the total intrinsic spin $S$, total orbital angular momentum $L$, both coupled to $J$, along with the intrinsic deformation. The deformation is informed by a pair of numbers $(\lambda\, \mu)$, where
$\lambda = N_{z} - N_{x}$ and $
\mu = N_{x} - N_{y}$ for a total of $N_{x} + N_{y} + N_{z}$ HO quanta distributed among the $x$, $y$, and $z$ directions (capped by $N_{\rm max}$).
The $(\lambda\, \mu)$ quantum numbers can be understood in the approximate mapping of the microscopic SU(3) basis states to a single rigid-rotor ellipsoid: 
The case where $N_{x} = N_{y} = N_{z}$, or equivalently $(\lambda\, \mu) = (0\, 0)$, describes a spherical configuration. A larger $N_{z}$ compared with $N_{x} = N_{y}$ (i.e., $\mu=0$ with $N_{z} > N_{x}$) indicates prolate quadrupole deformation; conversely, $(0\,\mu)$ indicates oblate quadrupole deformation.
Notably, a closed-shell configuration corresponds to $(0\, 0)$, which is spherical (or no deformation). 

The most dominant deformation, specified by $(\lambda\, \mu)$, for each state under consideration  along with the total intrinsic spin $S$ are listed in Table~\ref{tbl:level_theory}. States of the same major deformation and intrinsic spin can then be assigned as members of the same rotational band. Interestingly, the states below 8 MeV share the same triaxial deformation (in both $^8$B and $^8$Li) while coexistence of triaxial and oblate deformations is observed for the states above 8 MeV.

\begin{figure*}[!htb]
\includegraphics[width=1.\linewidth]{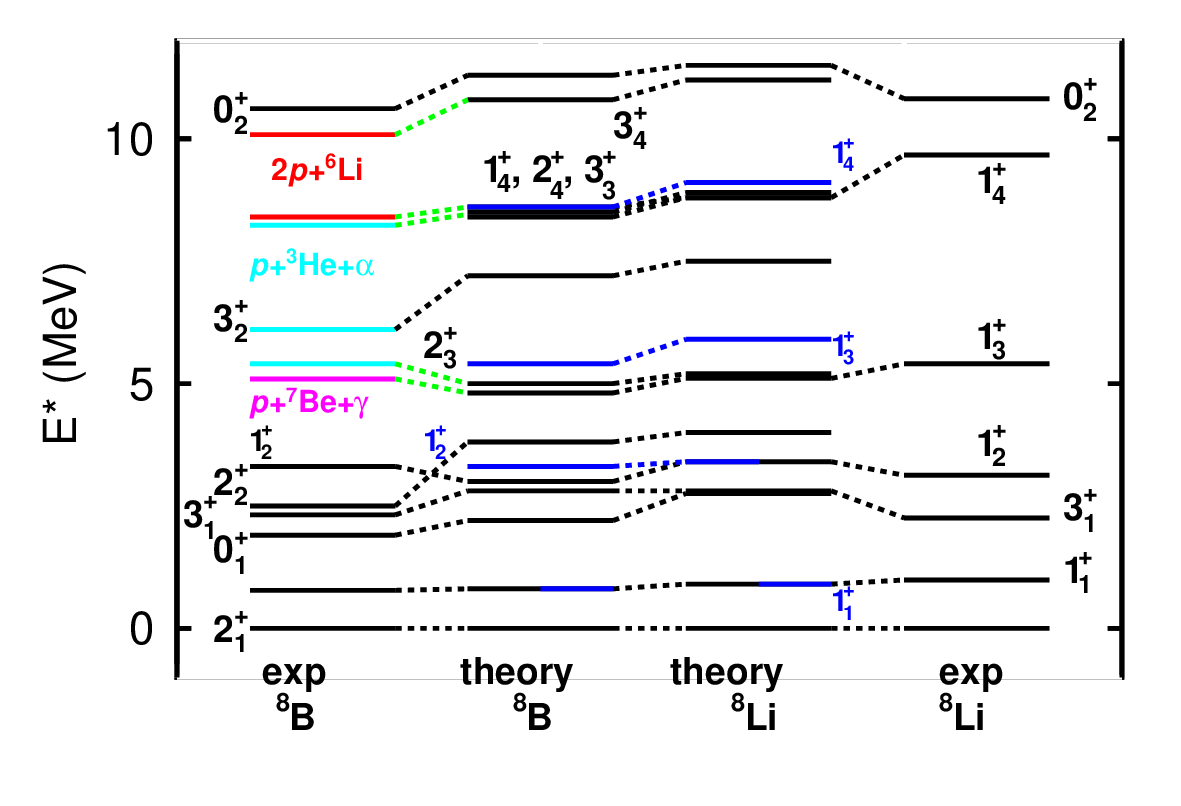}

\caption{
Comparison of levels schemes of the mirror nuclei $^8$B and $^8$Li from experiment and theory. The black experimental levels are from the ENSDF data base and from Ref.~\cite{Mitchel:2013} for the 0$^+_1$, 0$^+_2$, 2$^+_2$, and 1$^+_2$ states of $^8$B, while the magenta, red, and cyan colored levels are from the invariant-mass peaks observed in this work.  The magenta, cyan, and red colors correspond to the $p$+$^7$Be+$\gamma$, $p$+$^3$He+$\alpha$, and 2$p$+$^6$Li final exit channels respectively. Theoretical black levels are from the predictions of the SA-NCSM (third column of Table~\ref{tbl:level_theory}) while  the 1$^+$ levels in blue are from the GFMC
calculations with the AV18+IL7 interaction (fifth column of Table~\ref{tbl:level_theory}).  Connections between levels of the same $J^\pi$ 
are made with dotted lines. The green dotted lines indicate associations made in this work with the aid of the SA-NCSM predictions.
}
 \label{fig:level_theory}
\end{figure*}

For the SA-NCSM calculations in this work, we adopt the NNLO$_{\rm opt}$ chiral potential \cite{Ekstrom:2013}. We perform calculations for $\hbar\Omega=15,$ 20 and 25 MeV, and for model spaces up to $N_{\rm max}=12$. For each of the $\hbar\Omega$ values we perform extrapolations of the energies to obtain $N_\mathrm{max}\rightarrow\infty$ results by using a three-parameter exponential formula, similar to Ref. \cite{MarisVS09}: $E(N_\mathrm{max})=E(\infty)+a e^{-cN_\mathrm{max}}$. The extrapolated energies and their uncertainties are determined from $E(\infty)$ across the $\hbar\Omega$ range.

\begin{table}[t]
\caption{\label{tbl:level_theory} Predicted excitations energies of levels from the \textit{ab initio} SA-NCSM and the GFMC calculations. The latter values are only given for the 1$^+$ states and for two interactions. Also shown are the dominant total intrinsic spin and deformation, $S\,(\lambda\,\mu)$, from the SA-NCSM wave functions (see text for details).}

\begin{ruledtabular}
\begin{tabular}{clcccccc}

 State&   \multicolumn{3}{c}{SA-NCSM} &\multicolumn{2}{c}{GFMC } & \multicolumn{2}{c}{GFMC  }  \\
 &$S(\lambda\,\mu)$  & \multicolumn{2}{c}{NNLO$_{\rm opt}$} &\multicolumn{2}{c}{AV18+IL7} & \multicolumn{2}{c}{NV2+3-Ia} \\
    &&  $^8$B & $^8$Li & $^8$B & $^8$Li & $^8$B & $^8$Li \\
 \colrule
2$^+_1$	&	$	1 (2\, 1)\footnote{Bandhead of mostly triaxial $S=1$ rotational band.}	$	&	0	&	0	&	0	&	0	&	0	&	0\\	
1$^+_1$	&	$	1 (2\, 1)	$	&	0.8(1)	&	0.9(1)	&	0.8	&	0.9	&	1	&	0.7	\\
0$^+_1$	&	$	1 (2\, 1)	$	&	2.2(5)	&	2.8(5)	\\								
3$^+_1$	&	$	1 (2\, 1)	$	&	2.8(2)	&	2.8(2)	\\								
1$^+_2$	&	$	0 (2\, 1)	\footnote{Bandhead of mostly triaxial mixed $S=0$ rotational band. There is also a slightly smaller $S$=1 (2 1) contribution to this state.}$	&	3.0(4)	&	3.4(4)	&	3.3	&	3.4	&	3.3	&	3.6	\\
2$^+_2$	&	$	1 (2\, 1)	$	&	3.8(2)	&	4.0(2)	\\								
1$^+_3$	&	$	1 (2\, 1)	$	&	4.8(3)	&	5.1(3)	&	5.1	&	5.9	&	5.6	&	5.7	\\
2$^+_3$	&	$	0 (2\, 1)	$	&	5.0(1)	&	5.2(1)	\\								
3$^+_2$	&	$	1 (2\, 1)	$	&	7.2(2)	&	7.5(2)	\\								
1$^+_4$	&	$	1 (0\, 2)\footnote{Bandhead of mostly oblate $S=1$ rotational band.}	$	&	8.6(4)	&	8.8(4)	&	8.6	&	9.1	&	7.4	&	8.2	\\
2$^+_4$	&	$	1 (2\, 1)	$	&	8.4(4)	&	8.8(4)	\\								
3$^+_3$	&	$	0 (2\, 1)	$	&	8.5(4)	&	8.9(4)	\\								
3$^+_4$	&	$	1 (0\, 2)	$	&	10.8(2)	&	11.2(2)	\\								
0$^+_2$	&	$	0 (0\, 2)\footnote{Bandhead of mostly oblate $S=0$, $T=2$ rotational band.}	$	&	11.3(7)	&	11.5(7)	\\		\end{tabular}
\end{ruledtabular}
\end{table}

The SA-NCSM predicted level scheme for 0$^+$, 1$^+$, 2$^+$ and 3$^+$ states in $^8$B and its mirror $^8$Li are listed with uncertainties in Table~\ref{tbl:level_theory} and their centroids are
plotted in Fig.~\ref{fig:level_theory}. 
We restrict ourselves to these levels as they
can be produced following proton $p$-wave knockout from $^9$C. The predicted low-lying and previously known states, indicated by the black lines, find prospective experimental candidates within 500 keV, with larger deviation observed for $2^+_2$ and $3^+_2$. New states, indicated by red lines, also find candidates as discussed in Sec. \ref{sec:Dis}.

The predicted level schemes of the two mirrors are quite similar with no large Thomas-Ehrman shifts which could be produced by significant double occupancy of the proton second $s$ single-particle orbital. The largest predicted shift of 0.6 MeV is found for the 0$^+_1$ state (considering centroid energies only). State energies from the Greens function Monte Carlo (GFMC) approach \cite{Piarulli:Priv} using the AV18+IL7 phenomenological interaction \cite{Wiringa:1995,Pieper:2008} and the NV2+3-Ia chiral interaction \cite{Piarulli:2018} are also provided in Table~\ref{tbl:level_theory} for the 1$^+$ states. The energies from these two calculations are similar to the SA-NCSM values although the agreement is notably better with the AV18+IL7 interaction whose levels are shown as the blue lines in Fig.~\ref{fig:level_theory}. The energies of the low-lying excited states from SA-NCSM are also similar to the recent calculations from the no-core shell model with continuum for $^7$Be+p \cite{kravvaris2023ab}.

Spectroscopic factors and partial widths can be calculated from the \textit{ab initio} SA-NCSM wavefunctions following Ref.~\cite{SargsyanLSMD2023} for a single-nucleon fragment and Ref.~\cite{DreyfussLESBDD20} for a multi-nucleon fragment.

\begin{table*}[ht]
\caption{\label{tbl:SP} Spectroscopic factors calculated by the SA-NCSM for the overlap of 1$^+$, 2$^+$, and 3$^+$ states of interest in $^8$B with the  proton plus various $^7$Be states. Also shown are relevant estimated decay widths. These results are for $N_{\rm max}$=8 (10) model space for $p$ ($2p$) with $\hbar \Omega=15$~MeV HO frequency and using the NNLO$_{\rm opt}$ interaction. Width uncertainties are determined from the $^8$B theoretical energy uncertainties given in Table~\ref{tbl:level_theory}.}

\begin{ruledtabular}
\begin{tabular}{cccccccc}

 Channel&  1$^+_3$ & 1$^+_4$ & 2$^+_3$ & 2$^+_4$ & 3$^+_2$ & 3$^+_3$  &3$^+_4$ \\
 \colrule
 &\multicolumn{7}{c}{Spectroscopic factors}\\
$p$+$^7$Be(1/2$^-$) & 0.56 & 0.03 & 0.12 & 0.06 & $<$0.01 & $<$0.01 & $<$0.01 \\
$p$+$^7$Be(3/2$^-$) & 0.001 & 0.02 & 0.21 & 0.02 & 0.04 & 0.08 & $<$0.01\\
$p$+$^7$Be(5/2$^-_1$) & 0.62 & 0.01 & 0.64 & 1.08 & 0.25 & 0.81 & 0.12\\
$p$+$^7$Be(5/2$^-_2$) & 0.02 & 0.69 & 0.003 & 0.08 & 0.11 & 0.11 & 0.48 \\
$p$+$^7$Be(7/2$^-$) & $<$0.01 & 0.001 & 0.52 & $<$0.001 & 1.02 & 0.25 & 0.02\\
\colrule
 &\multicolumn{7}{c}{$\Gamma$ (MeV)}\\
 $p$+$^7$Be(5/2$^-_1$) && & & 0.8(4) & & 0.6(3) & 0.34(2)\\
 $p$+$^7$Be(5/2$^-_2$) && 0.3(2) & & & & & 1.2(1)\\
 Prompt $2p$+$^6$Li(1$^+_1$) & & 0.8(2) & & 0.04(2) & & 0.02(1) & 0.2(1) \\
 $^3$He($1/2^+_1$)+$^5$Li(3/2$^-_1$) & & 0.04(1) & & 0.2(1) & & 0.2(1) &  \\
 Total & & 1.1(3) & & 1.0(4) & & 0.8(3) & 1.7(2)
\end{tabular}
\end{ruledtabular}
\end{table*}

The results are reported in Table~\ref{tbl:SP} for $\hbar \Omega=15$~MeV HO frequency. The estimates do not include cluster normalization, which was found negligible for $p+^{6,7,8}$Li in Ref.~\cite{SargsyanLSMD2023}. For the proton channel, the SFs and partial widths are calculated from SA-NCSM overlaps $\braket{^8{\rm B}}{p+^7{\rm Be}}$ in $N_{\rm max}$=8 model spaces. The $2p+^6$Li and $^3$He$+^5$Li channels use SA-NCSM wavefunctions of $^8$B calculated in $N_{\rm max}$=10 model spaces, along with overlaps calculated for leading and subleading shape contributions within $^8$B and the decay products. The leading shape contribution, e.g., for $^8$B, comes from the $(\lambda \, \mu)$ deformation listed in Table~\ref{tbl:level_theory} and its vibrations (multiples of two-shell one-particle-one-hole excitations up through $N_{\rm max}$=10; see Refs.~\cite{DreyfussLESBDD20,LauneyMD_ARNPS21} for details). We note that the $2p+^6$Li width for the $3^+_3$ state and the $^3$He$+^5$Li width for the $1^+_4$ state are based on subleading contributions only. The widths for the prompt $2p+^6$Li decay include $s$, $p$, and $d$ waves, with $^1s_0$ and $^3p_{0,1,2}$ channels for the $pp$ system. The total widths are reported as a sum over all $^6$Li and $^3$He channels, including the $2p$ and $p\alpha$ sequential decays through the $^7$Be($5/2^-$) states. To estimate the widths for these sequential decays, we use branching ratios of 4\% ($p$) and 96\% ($\alpha$) for the $^7$Be($5/2^-_1$) state, and 99\% ($p$) and 1\% ($\alpha$) for the $^7$Be($5/2^-_2$) state, calculated from the $^7$Be SA-NCSM wavefunctions in $N_{\rm max}=8$ model spaces and using experimental threshold energies. These are consistent with results reported in the Gamow shell model~\cite{Fernandez:2023}. 
The $^7$Be($5/2^-_2$) alpha branching ratio is underestimated compared with 4.6\% reported in Ref.~\cite{McCray:1963}, but this small difference appears to be inconsequential to the total width estimates.

\section{Discussion}\label{sec:Dis}
In this section we associate the observed levels with those predicted by the \textit{ab initio} theories. In particular, we try to fix those that can be produced via a proton $p$-wave knockout from the $^9$C projectile. This process will populate positive-parity states with spins from 0 to 3. Weighted average energies and total widths from the different fits are listed in Table~\ref{tbl:weight} along with their associated $J^\pi$.

\begin{table*}[!ht]
\caption{ \label{tbl:weight} Weighted averages of the  energy and total widths from all the different fits 
of the newly observed levels in this work. Tentatively assigned spins are based on the comparisons to SA-NCSM predictions.  For the 3$^+_2$ state, the spin is constrained from the correlations. Intermediate states produced in the decay of these states are also listed. These intermediate states are those observed in the correlations, or when this was not possible, the strongest decay branch from the SA-NCSM calculations is given.}  

\begin{ruledtabular}
\begin{tabular}{ccccc}

J$^\pi$ & Final exit channel & Intermediate states(s) & $E^*$ & $\Gamma$ \\
        &                    &                        & (MeV) & (MeV) \\
\colrule        
(1$^+_3$) & $p$+$^7$Be + $\gamma$ & $^7$Be(1/2$^-$)\footnotemark[1] & 5.095(20) & 1.17(7) \\

(2$^{+}_3$) & $p$+$^3$He+$\alpha$ & $^7$Be(7/2$^-$)\footnotemark[2]& 5.34(11) & 0.40(36) \\  
3$^+_2$ &   $p$+$^{3}$He+$\alpha$ & $^7$Be(7/2$^-$)\footnotemark[1] & 6.082(21) & 0.81(2) \\  
($2^+_4,3^+_3$)        & $p$+$^3$He+$\alpha$ & $^5$Li(g.s.),$^7$Be(5/2$^-_1$) \footnotemark[1]  & 8.23(18) &1.30(33) \\
1$^+_4$ &   2$p$+$^6$Li   & & 8.40(4)  & 0.88(7) \\
  (3$^+_4$)      &    2$p$+$^6$Li & $^7$Be(5/2$^-_2$)\footnotemark[2] & 10.08(26) & 2.44(53) \\
\end{tabular}
\end{ruledtabular}
\footnotetext[1]{Observed}
\footnotetext[2]{Not observed, but suggested by the SA-NCSM calculations as the strongest decay mode.}
\end{table*}

Let us first start with the only new state for which the spin has been determined from the measured correlations, i.e., the second 3$^+$ state, observed in the $p$+$^7$Be(7/2$^-$)$\rightarrow p$+$^3$He+$\alpha$ decay path. 
The SA-NSCM prediction for the energy of this state is located at least 0.9~MeV higher than the experimental value, see Fig.~\ref{fig:level_theory}. 
However, the SA-NCSM predicts that this state has a very large spectroscopic factor for proton decay to the 7/2$^-$ state in $^7$Be, see Table~\ref{tbl:SP}, which is consistent with the measured correlations. This consistency gives more confidence in using the SA-NCSM predicted decay paths to make associations for the other observed levels. 

 Given the good energy agreement between theory and experiment for the other states within the low-lying rotational bands, the larger energy deviation seen for the 3$^+_2$ (and for 2$^+_2$) state, is likely due to the $S$=0 and 1 mixing in this and neighboring states, such as 2$^+_3$ (and for 1$^+_2$) (see Table~\ref{tbl:level_theory}). Since the degree of spin mixing is dictated by the underlying nuclear interaction, it will be interesting to study these nuclei with another chiral potential, which we leave for future work.

In this work, two previously unknown states at $E^*\approx$5~MeV have been observed and in this region of excitation energy there are only two predicted states (1$^+_3$ and 2$^+_3$), see Fig.~\ref{fig:level_theory}. 
The least energetic of the new states is observed in the $p$+$^7$Be(1/2$^-$) channel and is likely the 1$^+_3$ level, the lower of the two predicted states,  which has a significant predicted spectroscopic factor for this channel and far more decay phase-space than the alternative $p$+$^7$Be(5/2$^-_1$) channel which also has a significant spectroscopic factor, see Table~\ref{tbl:SP}. 
As theory predicts only one other positive-parity level in this energy region, the state observed in the $p$+$^3$He+$\alpha$ channel should be the 2$^+_3$ state. This 2$^+_3$ state is predicted to have significant spectroscopic strength for both the $p$+$^7$Be(7/2$^-$) and $p$+$^7$B(5/2$^-_1$) channels but only the former is energetically allowed. As the 7/2$^-$state in $^7$Be decays into the $^3$He+$\alpha$ channel, this assignment will produce the observed final state. This adds support to the spin assignments for both states in this energy region.  It is true however that the peak-to-background ratio for what we now assign as the 2$^+_3$ state was too small to allow us to verify if correlations between the decay products are consistent with the posited decay scenario and it is conceivable that another decay path, such as an initial $^3$He+$^5$Li split, is possible. One should also keep in mind that we have only considered positive-parity candidates.

Based on the SA-NCSM predictions shown in Fig.~\ref{fig:level_theory}, candidates levels  for the two observed peaks at $\approx$8.4~MeV associated with 2$p$+$^6$Li and $^3$He+$^5$Li$\rightarrow p$+$^{3}$He+$\alpha$ 
exit channels are the 1$^+_4$, 2$^+_4$ and the 3$^+_3$ states which have very similar predicted excitation energies and decay widths. In a companion paper \cite{CharityShortArxiv}, it is argued that the 2$p$+$^6$Li peak must be the 1$^+_4$ peak, with a total theoretical width [$\Gamma =1.1(3)$ MeV] consistent with the experimental value (see Tables~\ref{tbl:SP} and~\ref{tbl:weight}). The $p$+$^3$He+$\alpha$ peak may be a second decay branch of the same state as the extracted excitation energies and widths are similar. If that is the case, then from the relative yields of the two peaks, the branching ratio to the $p$+$^{3}$He+$\alpha$ channel would be 1.2(4) times that of the 2$p$ branch after correcting for detector efficiency.
However, the SA-NCSM calculations do not predict any significant decay branch to either the $p$+$^7$Be(5/2$^-_1$) or $^{3}$He+$^5$Li channels (Table~\ref{tbl:SP}) which, after sequential decay, will populate the $p$+$^3$He+$\alpha$ exit channel. 
Thus we consider the possibility that the 2$^+_4$ and the $3^+_3$ levels are the source of this yield.

Indeed, both of these states (2$^+_4$ and 3$^+_3$) have large spectroscopic factors  and significant widths for proton decay to the 5/2$^-_1$ state in $^7$Be (Table~\ref{tbl:SP}) which sequentially decays to the observed $p$+$^3$He+$\alpha$ final state. They also have much larger $^{3}$He+$^5$Li partial widths compared with that of the $1^+_4$ state. Theory predicts total decay widths of $\Gamma=1.0(4)$ MeV for the 2$^+_4$ state and $\Gamma=0.8(3)$ MeV for the 3$^+_3$ state which are consistent with the experimental result within the uncertainties of both experiment and theory.

Finally, a wide [$\Gamma$=2.44(53)~MeV] state at $E^*$=10.08~MeV was needed to fit the 2$p$+$^6$Li invariant-mass spectra. This proposed state is not well resolved in the invariant-mass spectra of Fig.~\ref{fig:Inv_2pLi6} and the excess yield above background could also be fit assuming more than one contributing level. It is interesting to explore whether a wide state is expected at this energy. The 2$^+_4$ and 3$^+_3$ levels are potential candidates but, as just discussed, they are expected to populate the $p$+$^3$He+$\alpha$ channel. The 3$^+_4$ state predicted at $E^*$=10.8~MeV is a clear candidate. This state is mostly an $L=2$ rotation of the 1$^+_4$ state but with the addition of non negligible $S=2$ configurations. Predicted partial decay widths, listed in Table~\ref{tbl:SP}, are dominated by sequential 2$p$ decay through the 5/2$^-_2$ state in $^7$Be which populates the observed exit channel. The predicted prompt 2$p$ decay width is suppressed compared with the 1$^+_4$ state as now $d$ wave emission is required and the significant $S$=2 component reduces the overlap with the ground state of $^6$Li. There is also a $p$+$^7$Be(5/2$^-_1$) branch which will give rise to a smaller yield in the $p$+$^3$He+$\alpha$ exit channel. The total predicted decay width of 1.7(2)~MeV is consistent with the  experimental value of 2.44(53)~MeV within the uncertainties of both. Thus it is plausible to assign some, or all, of the excess yield at 10.08~MeV to the 3$^+_4$ state.

\section{CONCLUSION}
New $^8$B resonances have been observed in the 2$p$+$^6$Li, $p$+$^3$He+$\alpha$, and $p$+$^7$Be exit channels following proton-knockout and projectile fragmentation reactions using fast $E/A\approx$69-MeV beams of $^9$C and $^{13}$O. The decay products were identified in the HiRA array and excitation energies and decay widths extracted using the invariant-mass method. Correlations between decay products have been investigated to identify the decay pathways to the three final exit channels. Examples of prompt 2$p$ decay, proton decay to particle-unstable $^7$Be states, and initial $^3$He+$^5$Li breakup, have been observed.

The structure of $^8$B and the mirror states in $^8$Li have also been studied theoretically in an \textit{ab initio} approach using the symmetry-adapter no-core shell model with continuum. Excitation energies and spectroscopic factors have been calculated for $^8$B states covering the experimental range of excitation energies with spin and parities consistent with proton $p$-wave knockout from $^9$C projectiles. This information was used to associate all of the observed resonances to theoretical counterparts. All the new levels can be associated with positive parity states expected to be produced following $p$-wave proton knockout from a $^9$C projectile. From the present data, there is no need to invoke any proton $s$-wave knockout producing negative-parity states.

 \begin{acknowledgments}

This work is supported by the U.S. Department of Energy, Office of Science, Office of Nuclear Physics under Awards No. DE-FG02-87ER-40316 and No. DE-SC0023532, and under the FRIB Theory Alliance Award No. DE-SC0013617. This work is also supported in part by the National Nuclear Security Administration through the Center for Excellence in Nuclear Training and University Based Research (CENTAUR) under Grant No. DE-NA-0004150. This work benefited from high performance computational resources provided by LSU (www.hpc.lsu.edu), the National Energy Research Scientific Computing Center (NERSC), a U.S. Department of Energy Office of Science User Facility at Lawrence Berkeley National Laboratory operated under Contract No. DE-AC02-05CH11231, as well as the Frontera computing project at the Texas Advanced Computing Center, made possible by U.S. National Science Foundation Award No. OAC-1818253.

  \end{acknowledgments}

 
\appendix

\section{Background for the 2$p$+$^6$Li events}
\label{sec:mix2pLi6}

A number of processes not associated with $^8$B decay can contribute to the detected yield of 2$p$+$^6$Li events. These other processes contribute to the background upon which the resonance peaks of interest sit.  For the $^9$C beam, a proton knockout is required to produce $^8$B states. An example of a background process is if the knocked-out proton is emitted inside the angular acceptance of the detector and is detected, but one of the protons from 2$p$ resonance decay is not. One can also consider the knockout of two protons forming a $^7$Be resonance that proton decays. If one of the two knocked-out protons and the resonance-decay proton are detected in coincidence with the $^6$Li residual, a background event is generated. Background is also generated if three protons are knocked out directly producing a $^6$Li residue and two of these knocked-out protons are detected.  There are more ways of producing background 2$p$+$^6$Li events in the fragmentation of the $^{13}$O. This background is not totally uncorrelated because there may be contributions from the decay of $^7$Be resonances present. A procedure to estimate the background in the invariant-mass spectra was presented in Ref.~\cite{Charity:2023} involving weighted event mixing. Herein we follow the procedure suggested in that work for the 2$p$ decay of $^{12}$O. 

\begin{figure}[!htb]
\includegraphics[width=1.\linewidth]{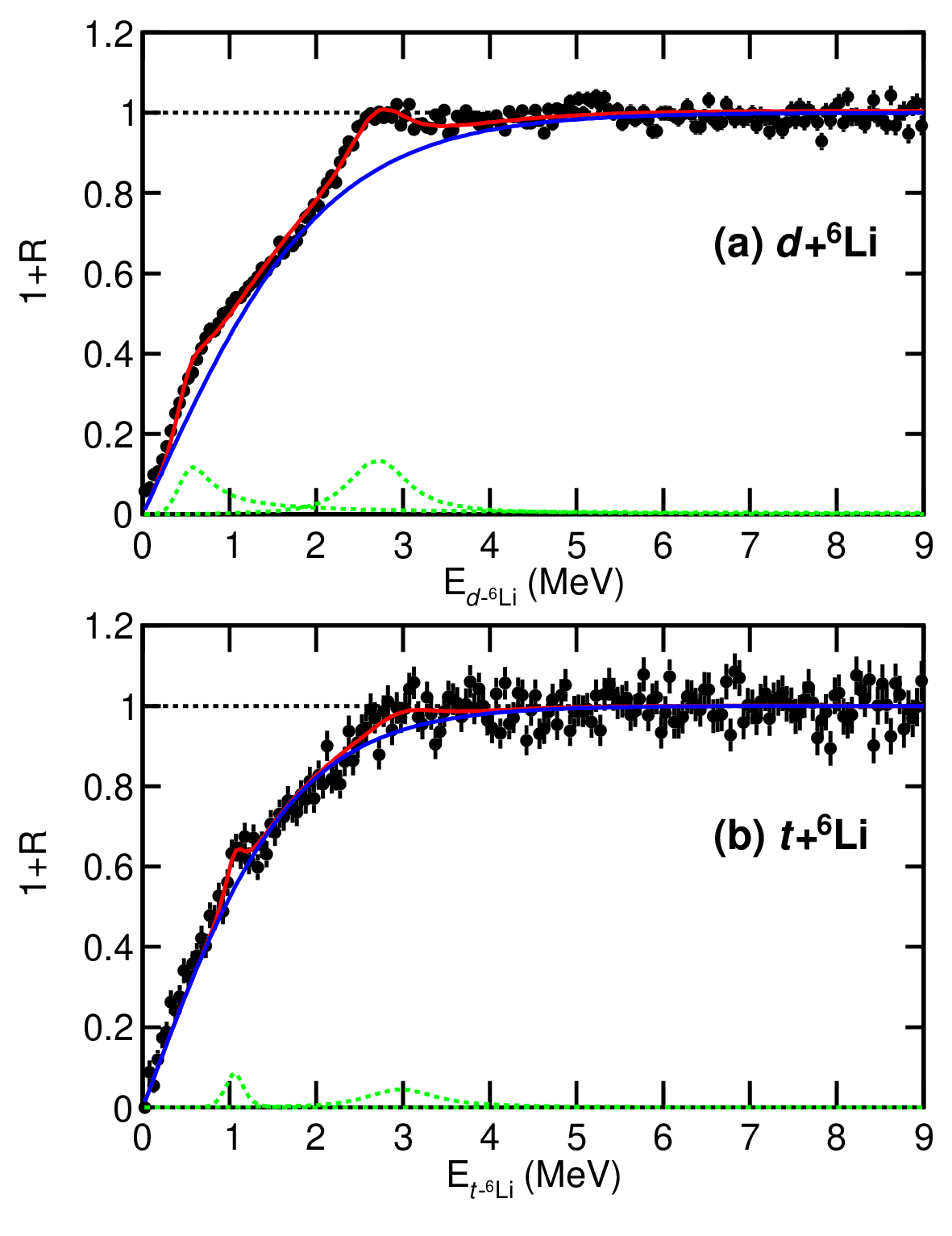}
\caption{The data points show the experimental correlation function for (a) $d$+$^6$Li and (b) $t$+$^6$Li events obtained from the $^{13}$O dataset. The red curves show fits to these with small contributions (dotted green curves) from resonances in $^8$Be and $^9$Be, respectively, in addition to the contribution from non-resonance products (blue curves).}

 \label{fig:backCorr}
\end{figure}

The background was constructed by mixing a $p$ from one $p$+$^6$Li event with $p$ and $^6$Li fragments from a second such event.  The added lone proton is like detecting a non-resonance proton produced in the knockout or fragmentation stage of the reaction.  In order to use these manufactured data, one has to weight the mixed events to suppress the background distribution at the smaller invariant masses. This suppression can be due to Coulomb correlations present in the fragmentation process. The weighting depends on the relative energy between the $p$ and $^6$Li fragments from the different events. In Ref.~\cite{Charity:2023} it was proposed that the weighting function should be  taken as the measured correlation function determined for $d$+core or $t$+core events because the Coulomb barriers are very similar to the $p$+core case. Furthermore, the $d$ and $t$ clusters produced in the fragmentation of a proton-rich projectile are generally not produced via resonant decay, but rather are created directly in the fragmentation process and so their correlation function would be free of resonances. However in this case, $d$+$^6$Li and $t$+$^6$Li events can also be produced from $^{8,9}$Be resonances. Indeed, in the $^{13}$O dataset, the correlation functions shown in Fig.~\ref{fig:backCorr}, obtained by dividing the experimental $E_{T}$ distributions by their mixed counterparts, show hints of resonances. The presence of the resonances is most pronounced for the $d$+$^6$Li correlation function, see Fig.~\ref{fig:backCorr}(a), but there are also suggestions of resonances in the $t$+$^6$Li events, see Fig.~\ref{fig:backCorr}(b). 

In order to remove the contribution from these resonances, leaving just the smoother Coulomb correlations, both correlation functions were fit with two resonances, dotted-green curves in Fig.~\ref{fig:backCorr}. The intrinsic lineshapes of these resonances were parametrized with the $R$-matrix formalism. These resonances sit upon the desired smooth weighting function of relative energy $E_{rel}$, solid-blue curves in Fig.~\ref{fig:backCorr}, parameterized by
\begin{equation}
    S(E_{rel}) = \frac{2}{1+\exp(-E_{rel}/c)} -1.  \label{eq:weight}
\end{equation}

For the $d$+$^6$Li data, the fitted peaks correspond to $^8$Be excited states at $E^*$=23.0(1) and 25.0(1)  MeV with intrinsic widths of  $\approx$0.5 and 0.62(6)~MeV,  respectively. The former is possibly a decay branch of the $E^*$=22.98(1)-MeV [$\Gamma$=0.23(5) MeV] state listed in the ENSDF data base \cite{ENSDF}.

For the $t$+$^6$Li data, there are indications of peaks at excitations energies 18.75 and 20.7 MeV in $^9$Be.  The fitted values of $c$ are similar, i.e.  1.05(4) and 0.86(4)~MeV for the $d$+$^6$Li and $t$+$^6$Li data, respectively. These are similar to the values of $c$=1.0(2) and 1.5(2)~MeV obtained from fitting the $d,t$+$^8$B and $d$+$^9$B correlations functions in Ref.~\cite{Charity:2023}, respectively,  obtained from the same dataset. 

For the $^9$C(2nd) dataset, $d,t$+$^6$Li events would require the pickup of neutrons from the target nucleus and are thus associated with a different reaction mechanism and hence their usefulness in determining the background in particle-removal reactions is unclear. In any case, these events have such low yields that the correlations functions cannot be determined with sufficient statistical accuracy to be employed.  Thus, in the fits to the 2$p$+$^6$Li invariant-mass spectra in Fig.~\ref{fig:Inv_2pLi6}, the same suppression factors were used in constructing the background for both datasets.
The parameter $c$ is allowed to span the range of the $d$+$^6$Li and $t$+$^6$Li values, i.e. 0.82 $< c < $ 1.05 with this uncertainty incorporated into the final uncertainty of the fit parameters.

\begin{figure}[!htb]
\includegraphics[width=1.\linewidth]{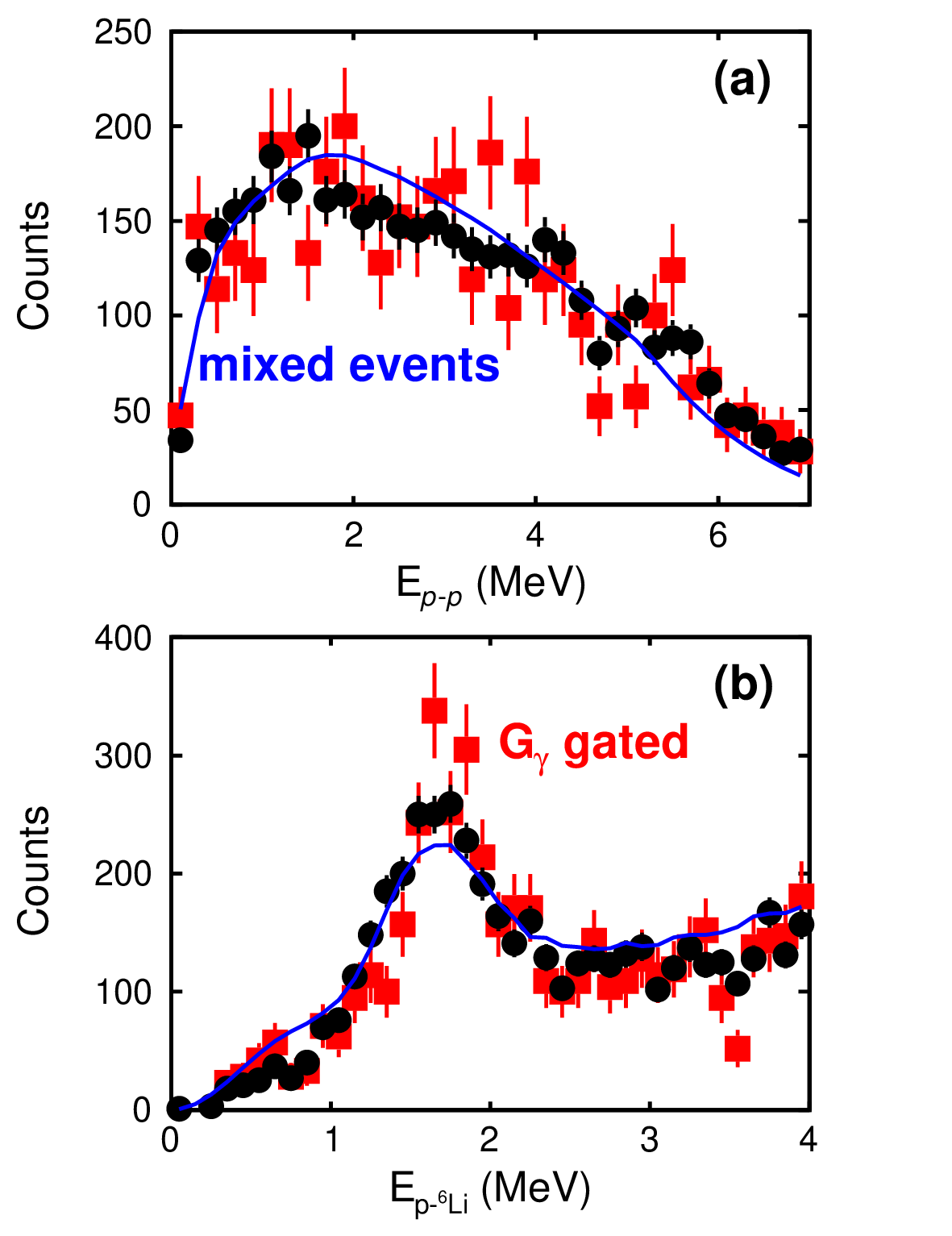}
\caption{Background correlations in the (a) $p-p$ and (b) $p$-core relative-energy distributions. Black circular data points show the correlations for the background region $B1$ in the distribution of $E^*_{n\gamma}$ from the 2$p$+$^6$Li exit channel in Fig.~\ref{fig:Inv_2pLi6} (also shown in Fig.~\ref{fig:gamma}). These data are from the $^9$C(2nd) dataset.  The blue curves are generated from the weighted mixed events and the square red data points are associated with real (unmixed) events with the extra condition of a coincident $\gamma$ ray in the $G2$ gate.}

 \label{fig:backCorr_2pLi6}
\end{figure}

In order to determine the 2$p$ correlations for the 8.4-MeV peak, one needs to subtract the significant background contribution under this peak.
For the $P1$ gate around this peak in Fig.~\ref{fig:Inv_2pLi6}, the contribution from the peak in the fits represents 64(4)\% and 60(4)\% of the total in the $^9$C(2nd) and $^{13}$O datasets, respectively. We have tried a number of ways of estimating the background, all of which give consistent outcomes. Two procedures, which are compared in  Fig.~\ref{fig:backCorr_2pLi6}, are using the weighted mixed events and using the $G2$-gated ($\gamma$ gated) events for which the 8.4-MeV peak is largely missing. We first looked at the region $B1$ in Fig.~\ref{fig:Inv_2pLi6} which is dominated by the background contributions in both datasets.  The circular back data points in Fig.~\ref{fig:backCorr_2pLi6} show the $p$-$p$ and $p$-$^6$Li relative energy distribution for these events from the $^9$C(2nd) dataset. The peak in the $p$+$^6$Li distribution is due to the 5/2$^-_2$ resonance in $^7$Be. The correlations obtained with the weighted mixed events, normalized to the same total yield (blue curves) are consistent with these. The square red data points, associated with the $G2$-gated data, are also consistent. The $\gamma$-gating thus does not appear to modify the background correlations and thus the correlations under the 8.4-MeV peak can be estimated from this subset of events as well.

\begin{figure}[!htb]
\includegraphics[width=1.\linewidth]{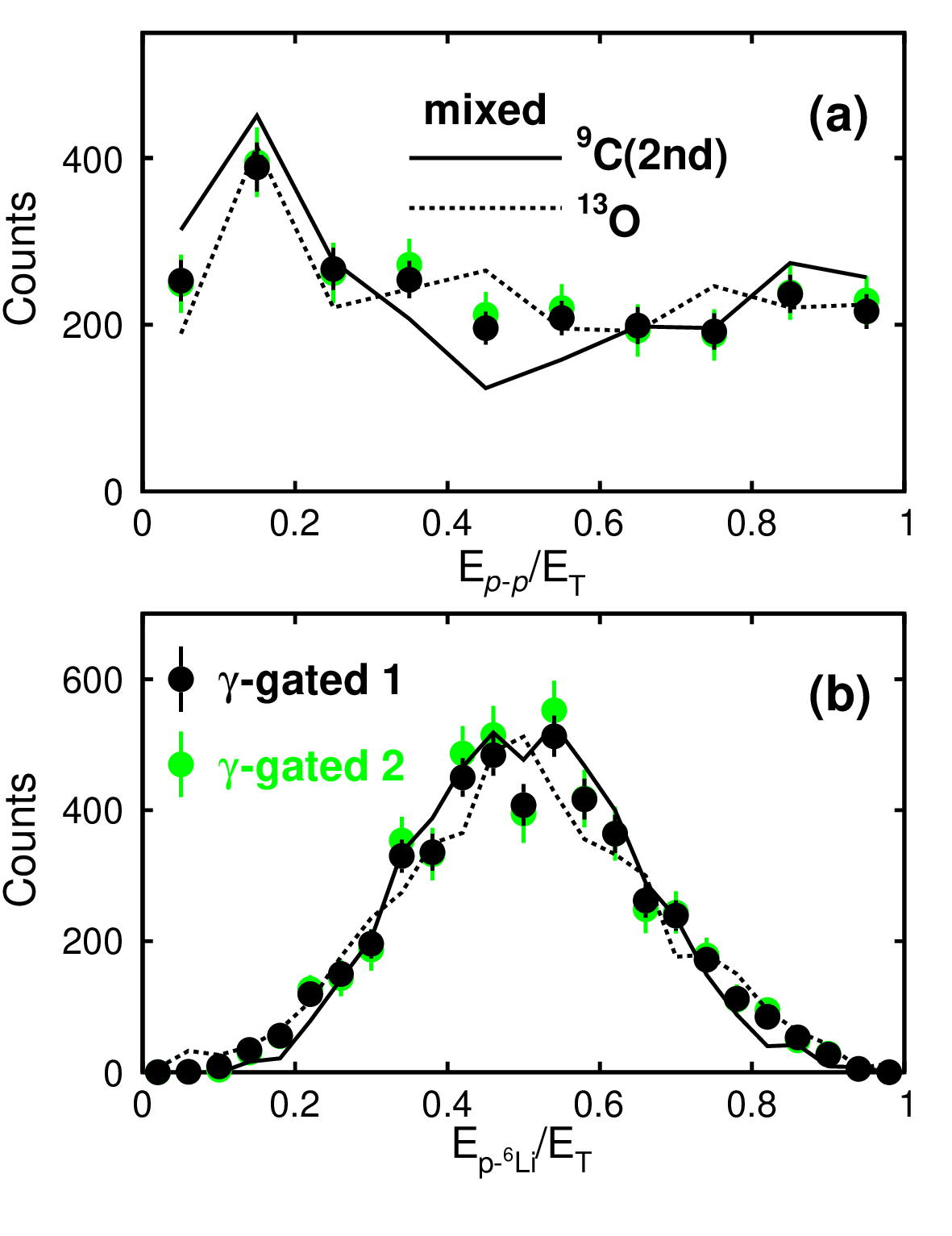}
\caption{ Comparison of background-subtracted correlations for the 8.4-MeV peak.
The correlations indicated by the solid and dashed curves are obtained from the $^9$C(2nd) and $^{13}$O datasets, respectively,  using the weighted mixed events to predict the background correlations (Appendix~\ref{sec:mix2pLi6}). The correlations given by the data points were obtained from the $\gamma$-gated events. The circular black data points (labeled $\gamma$-gated 1) were obtained assuming the $\gamma$-gated ($G2$) events have no contribution from the 8.4-MeV peak,  while the circular green data points (labeled $\gamma$-gated 2) were obtained when the small yield of this state obtained in the fit shown in Fig.~\ref{fig:gamma} was assumed. 
}
 \label{fig:Corr_2pLi6compare}
\end{figure}

Background-subtracted correlations for the 8.4-MeV state are compared in Fig.~\ref{fig:Corr_2pLi6compare}. The solid and dashed curves show the results from the $^9$C(2nd) and $^{13}$O datasets, respectively,  where the background correlations are taken from the respective weighted mixed events.  The results indicated by the black data points are from the $^9$C(2nd) datasets using the $\gamma$-gated events to provide the shape of the background, which is then normalized to the fitted background yield. The background estimate from this procedure assumes that there is no contribution of the 8.4-MeV state in the $\gamma$-gated spectrum in Fig.~\ref{fig:gamma}(b). The green data points arise when an allowance is made for a small yield of the 8.4-MeV state in this spectrum. This yield was obtained by a fit to the $G2$-gated spectrum, with the same peak and background components as in the fit to all events in Fig.~\ref{fig:Inv_2pLi6}(a). This fit is shown as the red curve in Fig.~\ref{fig:gamma}(b). There is excellent consistency among the four ways of obtaining the correlations, especially for the $p$-core relative-energy distribution in Fig.~\ref{fig:Corr_2pLi6compare}(b).

\section{Background for the $p$+$^3$He+$\alpha$ events}
\label{sec:mixpHe3a}
Given the width of the  $^7$Be(7/2$^-$) state is not that large ($\Gamma$=175~keV \cite{ENSDF}), the $p$+$^3$He+$\alpha$ events gated on the presence of this state (gate $P2$) are a good example of sequential decay. In constructing the background for these sequential events, we have mixed a proton from one gated $p$+$^3$He+$\alpha$ event with the identified  $^7$Be(7/2$^-$)$\rightarrow^{3}$He+$\alpha$ products from another such event. These mixed events are then weighted based on the $p$-$^7$Be relative energy using Eq.~\ref{eq:weight}. The uncertainty analysis considers the same range of the parameter $c$ as used in the 2$p$+$^6$Li events. 

The vetoed $p$+$^3$He+$\alpha$ events, i.e. without the $^7$Be(7/2$^-$) intermediate state,  represent a much more difficult task in estimating the background. There is no well-defined core and it is not clear how to mix the different fragments.  A few ways were tried and the variance in the results was included in the estimated uncertainty of the extracted peak parameters.  

We started by mixing a proton from one event with a $^{3}$He+$\alpha$ pair from another event, similar to our procedure for the $B2$-gated events.  In the $^{13}$O dataset, where there are no obvious peaks present in the vetoed events, no weighting of the mixed events is required in fitting the relevant spectrum.  Good fits of the $^9$C datasets were also obtained with an unweighted background.

As already noted, this channel is far more intense than the 2$p$+$^6$Li channel in the $^{13}$O dataset. 
It is possible that a substantial yield of these fragments comes from the sequential decay of $^{12}$N resonances produced following proton knockout from the projectile producing the $p$+$^3$He+2$\alpha$ channel where only one of the $\alpha$ particles is detected. In the subevents of this $p$+$^3$He+2$\alpha$ channel we find signatures of $^{11}$C, $^9$B, and $^8$Be resonances. These signatures are lost when only one of the two $\alpha$ particles is detected and, given the long lifetimes of these intermediate states, the Coulomb correlations would also be small. While this suggestion may potentially explain the absence of the Coulomb suppression for the $^{13}$O dataset, this mechanism will be absent for the two $^9$C datasets. We therefore include the possibly that the background from mixed events needs to the weighted in evaluating the uncertainty of the extracted resonance parameters. In this case, we weight the events of the form of Eq.~\ref{eq:weight} is based on the relative energy between the proton and the center of mass of the $^3$He and $\alpha$ particles spanning the same range of the $c$ parameter. This is probably an over estimation as we do not see evidence for other $^7$Be states present at significant levels.

We have also tried other ways of mixing events, i.e. mixing an $\alpha$ particle from one event with ${p}$+$^3$He fragments from another. In this case, suppression of events is required  even in the $^{13}$O dataset which is then applied to the two $^9$C datasets.  The backgrounds from the two ways of event mixing are quite similar, but given the small yield of the observed peaks in the $P2$-vetoed spectra, their differences dominate the uncertainty of the fitted peak parameters in Table~\ref{tbl:paHe3}, especially for the $E^*\approx$ 5.4-MeV state.

\begin{figure}[!htb]
\includegraphics[width=1.\linewidth]{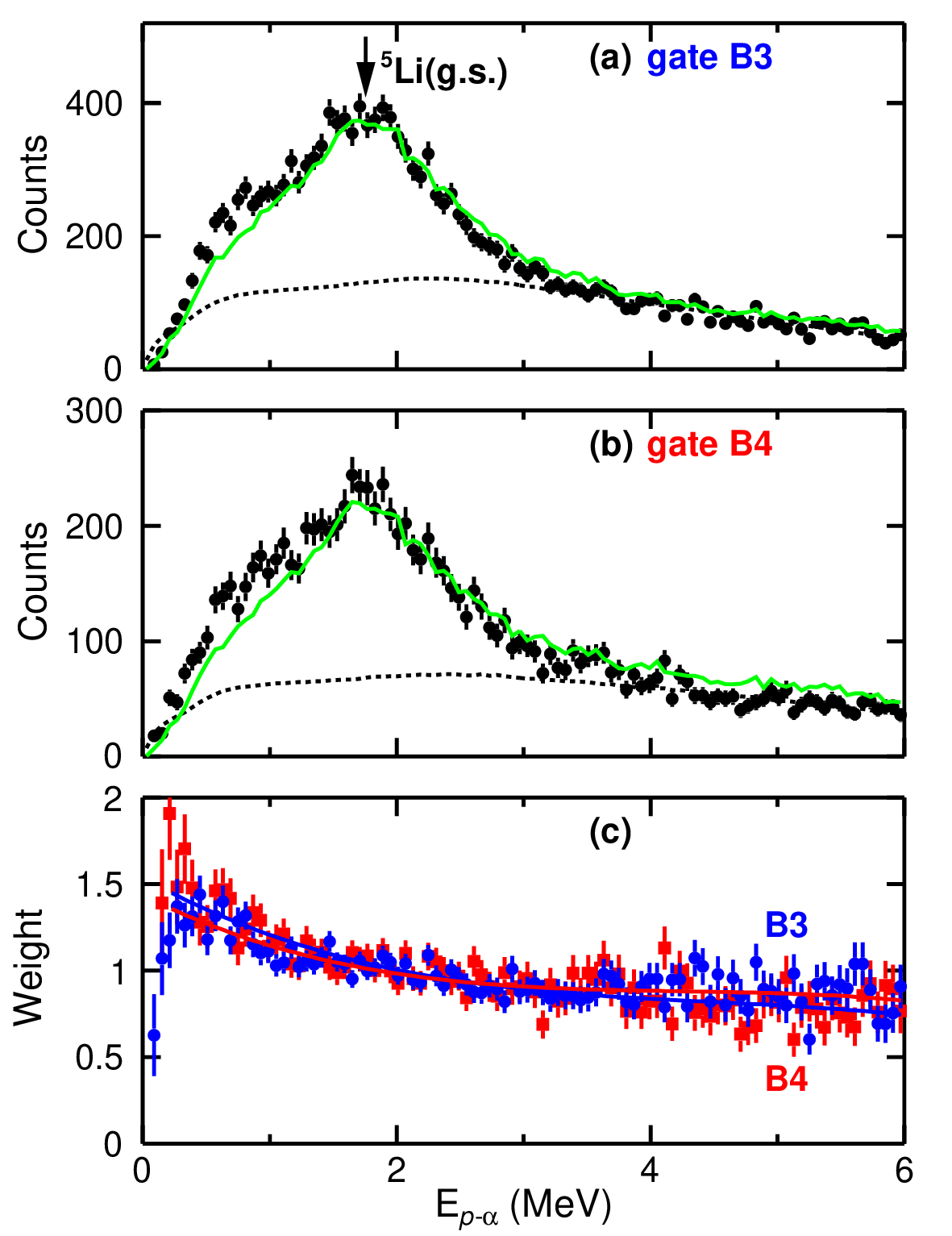}
\caption{Data points show the spectra of relative energy between the proton and $\alpha$-particle in $^7$Be(7/2$^-$)-vetoed $p$+$^3$He+$\alpha$ events from the $^9$C(1st) dataset for background gates (a) $B3$ and (b) $B4$ shown in Fig.~\ref{fig:Inv_pahe3}(c). The green curves are the results of the event mixing technique described in the text where a $^3$He fragment from one event is mixed with $p$+$\alpha$ subevents from another event.  The dotted curves represent 
the uncorrelated distributions where $\alpha$ and $p$ fragments from different events are mixed. (c) The ratio of the data to the green curves for both background gates. These ratios are fit with smooth polynomial curves. 
}
 \label{fig:Corr_pHe3a_back}
\end{figure}

While the different ways of mixing events produce similarly-shaped background distributions, the momentum correlations between the mixed fragments will depend very much on which fragments are from the same events and which are not. 
Background subtraction of the correlations for the 8.2-MeV state are thus not made from a single set of mixed events. The region of $E^*[^{8}B]$ above the 8.2-MeV peak is essentially all background in our fits [Fig.~\ref{fig:Inv_pahe3}(c)] and we use this region to extrapolate the background correlations to the region under the 8.2-MeV peak. As the background correlations are only presented as the relative-energy spectrum between each particles type, we only need to worry about the background in these quantities rather than in higher-order multi-dimensional correlations. 

The procedure we used for each of these one-dimensional backgrounds are equivalent and we only describe the method of extrapolating the backgrounds in the $p$-$\alpha$
relative-energy distribution.  For the other two relative-energy distributions, one can just replaces the roles of the different fragment types in the recipe described below.

The $p$-$\alpha$ background for the two gates $B3$ and $B4$ above the 8.2-MeV peak [Fig.~\ref{fig:Inv_pahe3}(c)] are shown in Figs.~\ref{fig:Corr_pHe3a_back}(a) and \ref{fig:Corr_pHe3a_back}(b), respectively. These data are from the $^9$C(1st) dataset with events vetoed by the $P2$ gate on $E^*$[$^{7}$Be]. The green curves are obtained from the $p$+$\alpha$ subevents  weighted in the following manner. Each
$p$+$\alpha$ subevent is mixed with a $^3$He fragment from other events multiple times. The weighting is determined from the probability that a mixed event has a reconstructed  $E^*$[$^8$B] value in the considered background gate. The weighted spectra are renormalized to the same total counts as their respective experimental distribution.  The weighting enhances the relative yield at high $E_{p-\alpha}$ values with increasing excitation energy. The green curves reproduce the experiment distributions remarkable well, with the only deficiency being a slightly reduced yield for the lower E$_{p-\alpha}$ values. 

The experimental distributions reveal strong correlations in the background.  To highlight these correlations, uncorrelated distributions are shown by the dotted curves. These uncorrelated distributions were obtained from mixing p+$^3$He ($^{3}$He+$\alpha$) subevents with $\alpha$ ($p$) fragments from another event, i.e.,  the $p$ and $\alpha$ particles come from different events. Both of these two mixing procedures give the same $E_{p-\alpha}$ distribution and they have been normalized to the tail regions of the experimental distributions in Figs.~\ref{fig:Corr_pHe3a_back}(a) and \ref{fig:Corr_pHe3a_back}(b). It is clear, that the background has a significant contribution from the $^5$Li resonance which are associated with the broad peak at $E_{p-\alpha}\approx$1.7 MeV [arrow in Fig.~\ref{fig:Corr_pHe3a_back}(a)]. There is also a low-energy shoulder on this resonance peak whose origin is unclear. 

Figure~\ref{fig:Corr_pHe3a_back}(c) shows the ratio of the experimental data to the green curves for both background gates. These ratios are almost identical for the two background regions which provides confidence that they can be applied in constructing the background for the region under the peak. The ratios were fit with polynomials and extrapolated into the $P3$ region under the peak in order to produce the background-subtracted distribution shown in Fig.~\ref{fig:corr_pahe3}. 
The largest contribution to the error bars for these distributions comes from the $\pm$4\% uncertainty in the magnitude of the background component.


\bibliography{extract_B8long}

\end{document}